\newcommand{\AP}[3]{Ann.\ Phys.\ {\bf #1},\ #2 (#3)}
\newcommand{\NPA}[3]{Nucl.\ Phys.\ {\bf A#1},\ #2 (#3)}
\newcommand{\NPB}[3]{Nucl.\ Phys.\ {\bf B#1},\ #2 (#3)}

\newcommand{\PLB}[3]{Phys.\ Lett.\ B\ {\bf #1},\ #2 (#3)}
\newcommand{\PR}[3]{Phys.\ Rep.\ {\bf #1},\ #2 (#3)}
\newcommand{\PRL}[3]{Phys.\ Rev.\ Lett.\ {\bf #1},\ #2 (#3)}

\newcommand{\PRC}[3]{Phys.\ Rev.\ C\ {\bf #1},\ #2 (#3)}
\newcommand{\PRD}[3]{Phys.\ Rev.\ D\ {\bf #1},\ #2 (#3)}
\newcommand{\JPG}[3]{J.\ Phys.\ G\ {\bf #1},\ #2 (#3)}

\newcommand{\ZPC}[3]{Z.\ Phys.\ C\ {\bf #1},\ #2 (#3)}

\newcommand{\EPJA}[3]{Eur.\ Phys.\ J.\ A\ {\bf #1},\ #2 (#3)}
\newcommand{\PTP}[3]{Prog.\ Theo.\ Phys.\ {\bf #1},\ #2 (#3)}

\newcommand{\diracslash}[1]{#1\llap{/\kern2pt}}

\newcommand{\be}{\begin{equation}}
\newcommand{\ee}{\end{equation}}
\newcommand{\bea}{\begin{eqnarray}}
\newcommand{\eea}{\end{eqnarray}}
\newcommand{\ba}[1]{\begin{array}{#1}}
\newcommand{\ea}{\end{array}}

\documentclass[prd,aps,floats,nofootinbib,tightenlines,showpacs]{revtex4-1}
\usepackage{epsfig,graphicx,pstricks}
\usepackage{psfrag}
\usepackage{color}
\usepackage{amsmath}
\usepackage{amsfonts}
\usepackage{amssymb}
\usepackage{textcomp}
\usepackage{multirow}

\begin{document}

\title {Vacuum structure and chiral symmetry breaking in strong magnetic fields for hot and dense quark matter
}
\author{Bhaswar Chatterjee}
\email{bhaswar@prl.res.in}
\affiliation{Theory Division, Physical Research Laboratory,
Navrangpura, Ahmedabad 380 009, India}
\author{Hiranmaya Mishra}
\email{hm@prl.res.in}
\affiliation{Theory Division, Physical Research Laboratory,
Navrangpura, Ahmedabad 380 009, India}
\author{Amruta Mishra}
\email{amruta@physics.iitd.ac.in}
\affiliation{Department of Physics, Indian Institute of Technology, New 
Delhi-110016,India}
%\affiliation{Frankfurt Institute for Advanced Studies,
%Universit\"at Frankfurt, D-60438 Frankfurt, Germany}
%\email{mishra@th.physik.uni-frankfurt.de}

\date{\today} 

\def\be{\begin{equation}}
\def\ee{\end{equation}}
\def\bearr{\begin{eqnarray}}
\def\eearr{\end{eqnarray}}
\def\zbf#1{{\bf {#1}}}
\def\bfm#1{\mbox{\boldmath $#1$}}
\def\hf{\frac{1}{2}}
\def\sl{\hspace{-0.15cm}/}
\def\omit#1{_{\!\rlap{$\scriptscriptstyle \backslash$}
{\scriptscriptstyle #1}}}
\def\vec#1{\mathchoice
        {\mbox{\boldmath $#1$}}
        {\mbox{\boldmath $#1$}}
        {\mbox{\boldmath $\scriptstyle #1$}}
        {\mbox{\boldmath $\scriptscriptstyle #1$}}
}

\begin{abstract}
We investigate chiral symmetry breaking in strong magnetic fields
at finite temperature and densities in a 3 flavor Nambu Jona Lasinio (NJL)
model including the Kobayashi Maskawa t-Hooft (KMT) determinant term,
using an explicit structure for the ground state in terms of
quark antiquark condensates.
The mass gap equations are solved self consistently and are used to
compute the thermodynamic potential. We also derive the equation of state
for strange quark matter in the presence of strong magnetic fields
which could be relevant for proto-neutron stars.
\end{abstract}

\pacs{12.38.Mh, 11.30.Qc, 71.27.+a, 12.38-t}

\maketitle

 \section{Introduction}
 The structure of QCD vacuum and its modification under
extreme environment has been a major theoretical and experimental
challenge in current physics \cite{review}. In particular, it is interesting to
study the modification of the structure of ground state at  high 
temperature and/or high baryon densities
as related to the nonperturbative aspects of QCD. This is important not only from a
theoretical point of view, but also for many applications to problems of
quark gluon plasma that could be  copiously produced in
relativistic heavy ion collisions as well as for the ultra dense cold
nuclear/quark matter which could be present in the interior of compact stellar objects
like neutron stars.
In addition to hot and dense QCD, the effect of strong magnetic field
on QCD vacuum structure has attracted recent attention.This is motivated
by the possibility of creating ultra strong magnetic fields in non central collisions
at RHIC and LHC. The strengths of the magnetic fields are estimated to be of hadronic
scale \cite{larrywarringa,skokov} of the order of $eB\sim 2 m_\pi^2$ 
($m_\pi^2\simeq 10^{18}$ Gauss) at RHIC, to
about $eB\sim 15 m_\pi^2$ at LHC \cite{skokov}. 

There have been recent calculations both analytic as well as with lattice simulations,
which indicate that QCD phase digram is affected by 
strong magnetic fields \cite{dima,maglat,fraga}. One of the interesting findings
has been the  chiral magnetic effect.
Here an electric current of quarks along the magnetic field axis is generated
if the densities of left and right handed quarks are not equal. At high temperatures
and in presence of magnetic field such a current can be produced locally.
The phase structure of dense matter in presence of magnetic field
along with a non zero chiral density has recently been investigated for two flavor
PNJL model for high temperatures relevant for RHIC and LHC \cite{fukushimaplb}.
There have also been many investigations to look into the vacuum structure of QCD
and it has been recognized that the strong magnetic field acts as a catalyser
of chiral symmetry breaking \cite{igormag,miranski,klimenko,boomsma}. 

In the context of cold dense matter, compact stars can be strongly
magnetized. Neutron star observations indicate the magnetic field to be of the order
of $10^{12}$-$10^{13}$ Gauss at the surface of ordinary pulsars  \cite{somenath}.
Further, the magnetars which are strongly magnetized
neutron stars, may have  even stronger
magnetic fields of the order of 
$10^{15}-10^{16}$ Gauss \cite{dunc,duncc,dunccc,duncccc,lat,broder,lai}.
Physical upper limit 
on the magnetic field in a gravitationally bound star is $10^{18}$ Gauss
that is obtained by comparing the magnetic and gravitational energies using
virial theorem \cite{somenath}. This limit could be higher for self bound
objects like quark stars \cite{ferrer3}. Since the magnetic field strengths
are of the order of QCD scale, this can affect both the thermodynamic as well
as hydrodynamics of such magnetized matter \cite{armendirk}.
The effects of magnetic field on the equation of state have been recently studied
in Nambu Jona Lasinio model at zero temperature for three flavors and
the equation of state  has been computed for the cold quark matter \cite{providencia}.
 It will be interesting to consider the finite temperature effects on such equation 
of state, which could be of relevance for proto-neutron stars.

We had earlier considered a variational approach to study chiral symmetry breaking
as well as color superconductivity in hot and dense matter with an explicit
structure for the `ground state' \cite{hmspmnjl,hmparikh,hmam}. The calculations 
were carried out
within NJL model with minimization of free energy density to decide which
 condensate will exist at what density and/or temperature. A nice feature of the 
approach is that the four component quark field operator in the 
chiral symmetry broken phase gets determined from the vacuum structure. 
In the present work, we aim to investigate how the vacuum structure in the 
context of chiral symmetry breaking gets modified in the presence of magnetic field.

 We organize the paper as follows. In  section II, we discuss the ansatz state
with quark antiquark pairs in the presence of a magnetic field.
We then generalize such a state to include the effects of temperature 
and density. In section III, we consider the 3 flavor NJL model along with the so called
KMT term--the six fermion determinant interaction term which breaks U(1) 
axial symmetry as in QCD. We use this Hamiltonian and calculate its 
expectation value with respect to  the ansatz state
to compute the energy density as well the thermodynamic potential 
for this system. We minimize the thermodynamic potential to determine
the the ansatz functions and the resulting mass gap equations.
We discuss the results of the present investigation in section IV. 
Finally we summarize
and conclude in section V. For the sake of completeness we derive the spinors
in the presence of a magnetic field and some of their properties, which are presented
 in the appendix.

\section{ The ansatz for the ground state}
To make the notations clear, we first write down the  field operator
expansion in the momentum space in the presence of a constant magnetic field $\zbf B$
in the $z-$ direction for a quark with a current mass $m$ and electric charge $q$.
 We choose the gauge such that the electromagnetic vector potential
is given as $A_\mu(\vec x)=(0,0,Bx,0)$. The Dirac field operator for a particle
is given as \cite{kausik}

\be
 \psi(\zbf x) = \sum_n\sum_r\frac{1}{2\pi}\int{d\zbf p_{\omit x}\left[q_r(n,\zbf p_{\omit x})
U_r(x,\zbf p_{\omit x},n) + \tilde q_r(n,-\zbf p_{\omit x})V_r(x,-\zbf p_{\omit x},n)\right]
e^{i\zbf p_{\omit x}\cdot\zbf x_{\omit x}}}.
\label{psiex}
\ee
The sum over $n$ in the above expansion runs from 0 to infinity.
In the above, 
$\zbf p_{\omit x}\equiv (p_y,p_z)$, and, $r=\pm 1$ denotes
the up and down spins.
We have suppressed 
the color and flavor indices of the quark field operators. 
The quark annihilation and antiquark creation operators,
$q_r$ and $\tilde q_r$,
 respectively, satisfy the quantum algebra
\begin{equation}
 \lbrace q_r(n,\zbf p_{\omit x}),q_{r^\prime}^\dag(n^\prime,\zbf p_{\omit x}^\prime)
\rbrace =
\lbrace \tilde q_r(n,\zbf p_{\omit x}),\tilde q_{r^\prime}^\dag(n^\prime,\zbf p_{\omit x}^\prime)\rbrace =
\delta_{rr^\prime}\delta_{nn^\prime}\delta(\zbf p_{\omit x}-\zbf p_{\omit x}^\prime).
\label{acom}
\end{equation}

In the above, $U_r$ and $V_r$ are the four component spinors
for the quarks and antiquarks respectively. The explicit forms of the spinors
for the fermions with mass $m$ and electric charge $q$ are given by
\begin{subequations}
\begin{eqnarray}
U_{\uparrow }(x,\vec p_{\omit x},n)&=&\frac{1}{\sqrt{2\epsilon_n(\epsilon_n+m)}}
\left(\begin{array}{c}
(\epsilon_n+m)\left(\Theta(q)I_n + \Theta(-q)I_{n-1}\right)\\
0\\
p_z\left(\Theta(q)I_n+\Theta(-q)I_{n-1}\right)\\
-i\sqrt{2n|q|B}\left(\Theta(q)I_{n-1}+\Theta(-q)I_{n}\right)\\
\end{array}\right) \\
 U_{\downarrow}(x,\vec p_{\omit x},n)& =& \frac{1}{\sqrt{2\epsilon_n(\epsilon_n+m)}}
\left(\begin{array}{c}
0 \\ 
(\epsilon_n+m)\left(\Theta(q)I_{n-1}+\Theta(-q)I_n\right) \\
i\sqrt{2n|q|B}\left(\Theta(q)I_n-\Theta(-q)I_{n-1}\right)
\\
-p_z\left(\Theta(q)I_n-\Theta(-q)I_{n-1}\right)\\
\end{array}\right) \\
V_{\uparrow}(x,-\zbf p_{\omit x},n) &=& \frac{1}{\sqrt{2\epsilon_n(\epsilon_n+m)}}
\left(\begin{array}{c}
\sqrt{2n|q|B}
\left(\Theta(q)I_n-\Theta(-q)I_{n-1}\right)
\\ 
ip_z
\left(\Theta(q)I_{n-1}+\Theta(-q)I_{n}\right)
\\
0\\
i(\epsilon_n+m)
\left(\Theta(q)I_{n-1}+\Theta(-q)I_{n}\right)
\\
\end{array}\right)\\
V_{\downarrow}(x,-\vec p_{\omit x},n) &=& \frac{1}{\sqrt{2\epsilon_n(\epsilon_n+m)}}
\left(\begin{array}{c}
ip_z
\left(\Theta(q)I_{n}+\Theta(-q)I_{n-1}\right)
\\ 
\sqrt{2n|q|B}
\left(\Theta(q)I_{n-1}-\Theta(-q)I_{n-1}\right)
 \\
-i(\epsilon_n+m)
\left(\Theta(q)I_{n}+\Theta(-q)I_{n-1}\right)
\\
0\\
\end{array}\right).
\end{eqnarray}
\label{UVs}
\end{subequations}
In the above, the energy of the n-th Landau level is given as
 $\epsilon_n=\sqrt{m_i^2+p_z^2+2n|q_i|B}
\equiv\sqrt{m_i^2+|\zbf p_i^2|}$. In Eq.s (\ref{UVs}), the functions
$I_n`$s (with $n\ge 0$) are functions of $\xi=|q_iB|(x-p_y/|q_iB|)$ and are given as
\be
I_n(\xi)=c_n\exp\left(-\frac{\xi^2}{2}\right)H_n(\xi)
\label{inxi}
\ee
where, $H_n(\xi)$ is the Hermite polynomial of
 the nth order and $I_{-1}=0$.  The normalization
constant $c_n$ is given by
\begin{equation*}
 c_n = \sqrt{\frac{\sqrt{|q|B}}{n!2^n\sqrt{\pi}}}
\end{equation*}
The functions $I_n(\xi)$ satisfy the orthonormality condition
\begin{equation}
 \int{d\xi I_n(\xi)I_m(\xi)} = \sqrt{|q|B}\delta_{n,m}
\label{orthoI}
\end{equation}
so that the spinors are properly normalized.
The detailed derivation of these spinors and some of their
properties are presented in the appendix.

With  the field operators now defined in terms of the annihilation and the 
creation operators
in presence of a constant magnetic field,
we now write down an ansatz for the ground state taken as a squeezed
coherent state involving quark and antiquarks pairs as
\cite{amhm5,hmam,hmparikh} 
\begin{equation} 
|\Omega\rangle= {\cal U}_Q|0\rangle.
\label{u0}
\end{equation} 
Here, ${\cal U}_Q$ is an  unitary operator  which creates
quark--antiquark  pairs from the vacuum $|0\rangle$. Explicitly,
the operator, ${\cal U}_Q$ is given as
\begin{equation}
{\cal U}_Q=\exp\left(
\sum_{n=0}^{\infty} \int{d\vec p_{\omit x} 
{q_r^i}^\dag(n,\vec p_{\omit x})a_{r,s}^i(n,p_z)f^i
(n,\vec p_{\omit x})
\tilde q_s^i(n,-\vec p_{\omit x})} -h.c.\right)
\label{ansatz}
\end{equation}
where, we have explicitly retained the flavor index $i$ for the quark field operators.
In the above ansatz for the ground state, $f^i(n,p_z)$ is a real function 
describing the quark antiquark
condensates related to the vacuum realignment for chiral symmetry breaking. 
In the above equation, the spin dependent structure  $a_{r,s}^i$ is given by 
\begin{equation}
 a_{r,s}^i=\frac{1}{|\zbf p_i|}\left[-\sqrt{2n|q_i|B}\delta_{r,s}-ip_z\delta_{r,-s}\right]
\end{equation}
with $|\zbf p_i| = \sqrt{p_z^2+2n|q_i|B} $ denoting the magnitude 
of the three momentum of the quark/antiquark of $i$-th flavor 
(with electric charge $q_i$) in presence of a magnetic field.
It is easy to show 
that, $a a^\dagger=I$, where
$I$ is an identity matrix in two dimensions. The ansatz functions $f_i(n,p_z)$
are determined from the minimization of thermodynamic potential. This particular
ansatz of Eq.(\ref{ansatz}) is a direct generalization of the ansatz 
considered earlier \cite{hmspmnjl,hmam}, to include the effects of magnetic field.
Clearly, a nontrivial $f_i(n,p_z)$ breaks the chiral
symmetry. Summation over three colors is understood in the
exponent of ${\cal U}_Q$ in Eq. (\ref{ansatz}). 

It is easy to show that the transformation of the ground state as in Eq.(\ref{u0})
is a Bogoliubov transformation. With the ground state transforming as Eq.(\ref{u0}),
any  operator $O$ in the $|0\rangle$ basis  transforms as 
\be
O'={\cal U}_Q O{\cal U}_Q ^\dag
\ee
and, in particular, the creation and the annihilation operators of Eq.(\ref{psiex})
transform as
\begin{eqnarray}
\left[\begin{array}{c}
q_r^\prime(n,\zbf p_{\omit x})\\
\tilde q_s^\prime(n,-\zbf p_{\omit x})
\end{array}\right] &=&
U_Q
\left[\begin{array}{c}
q_r(n,\zbf p_{\omit x})\\
\tilde q_s(n,-\zbf p_{\omit x})
\end{array}\right] 
U_Q^\dag \nonumber\\
&=&
\left[\begin{array}{cc}
\cos |f| & -a_{r,s}\sin |f|\\
a^\dag_{s,r}\sin |f| & \cos |f|
\end{array}\right]
\left[\begin{array}{c}
q_r(n,\zbf p_{\omit x})\\
\tilde q_s(n,-\zbf p_{\omit x})
\end{array}\right], \\
\label{bogtrans}\nonumber
\end{eqnarray}
 which is a Bogoliubov transformation with the transformed `primed' operators satisfying
 the same anti-commutation relations as the `unprimed' ones as in Eq.(\ref{acom}).
Using the transformation Eq. (\ref{bogtrans}),
we can expand the quark field operator $\psi(\zbf x)$ in terms of the primed operators
given as, 
%which will be useful later while calculating the expectation values.
\be
 \psi(\zbf x) = \sum_n\sum_r\frac{1}{2\pi}\int{d\zbf p_{\omit x}
\left[q_r^\prime(n,\zbf p_{\omit x})
U_r^\prime(x,n,\zbf p_{\omit x}) + \tilde q_r^\prime(n,-\zbf p_{\omit x})V_r^\prime(x,n,-\zbf p_{\omit x})\right]
e^{i\zbf p_{\omit x}\cdot\zbf x_{\omit x}}},
\label{psip}
\ee
with $q^\prime|\Omega\rangle=0={\tilde q}^{\prime\dag}|\Omega\rangle$. In the above, we 
have suppressed the flavor and color indices.
 It is easy to see that the `primed' spinors are give as
\begin{subequations}
\begin{eqnarray}
U_r^\prime(x,n,p_{\omit x}) = \cos |f| U_r(x,n,p_{\omit x}) - a_{r,s}^\dag\sin |f| V_s(x,n,-p_{\omit x}) \\
V_r^\prime(x,n,-p_{\omit x}) = 
\cos |f| V_r(x,n,-p_{\omit x}) + a_{s,r}\sin |f| U_s(x,n,p_{\omit x}).
\end{eqnarray}
\end{subequations}
Explicit calculation, e.g. for positive charges yield
 the following forms of the primed spinors:

\begin{subequations}
\begin{eqnarray}
 U_\uparrow^\prime(\vec p_{\omit x},n) &=& \frac{1}{\sqrt{2\epsilon_n(\epsilon_n+m)}}
\left[\begin{array}{c}
a_1 I_n\\
0\\
a_2 p_z I_n\\
-ia_2\sqrt{2n|q|B}I_{n-1}\\
\end{array}\right]  \\
 U_\downarrow^\prime(\vec p_{\omit x},n) &=& \frac{1}{\sqrt{2\epsilon_n(\epsilon_n+m)}}
\left[\begin{array}{c}
0\\
a_1 I_{n-1}\\
ia_2 \sqrt{2n|q|B}I_n\\
-a_2 p_z I_{n-1}\\
\end{array}\right] \\
V_\uparrow^\prime(-\vec p_{\omit x},n) &=& \frac{1}{\sqrt{2\epsilon_n(\epsilon_n+m)}}
\left[\begin{array}{c}
a_2 \sqrt{2n|q|B}I_n\\
ia_2 p_z I_{n-1}\\
0\\
ia_1 I_{n-1}\\
\end{array}\right] \\
V_\downarrow^\prime(-\vec p_{\omit x},n) &=& \frac{1}{\sqrt{2\epsilon_n(\epsilon_n+m)}}
\left[\begin{array}{c}
ia_2 p_z I_n\\
a_2\sqrt{2n|q|B}I_{n-1} \\
-ia_1 I_n\\
0\\
\end{array}\right], 
\label{primed+vespinors}
\end{eqnarray}
\end{subequations}
where the functions, $a_1$ and $a_2$, are given in terms of the
condensate function $f(p_z,n)$ as
\begin{eqnarray}
a_1 &=& (\epsilon_n+m)\cos |f(n,p_z)| + |\zbf p_i|\sin |f(n,p_z)| \\
a_2 &=& \cos |f(n,p_z)| - \frac{\epsilon_n+m}{|\zbf p_i|}\sin |f(n,p_z)|,
\end{eqnarray}

Let us note that with Eq.(\ref{psip}),
the four component quark field operator gets defined in terms of the
vacuum structure for chiral symmetry breaking given 
through Eq.(\ref{u0}) and Eq.(\ref{ansatz}) \cite{amspm,spmindianj}.

To include the effects of temperature and density we next write
 down the state at finite temperature and density 
$|\Omega(\beta,\mu)\rangle$  through
a thermal Bogoliubov transformation over the state $|\Omega\rangle$ 
using the thermo field dynamics (TFD) method as described in Ref.s \cite{tfd,amph4}.
This is particularly useful while dealing with operators and expectation values.
We write the thermal state as
\begin{equation} 
|\Omega(\beta,\mu)\rangle={\cal U}_{\beta,\mu}|\Omega\rangle={\cal U}_{\beta,\mu}
{\cal U}_Q |0\rangle,
\label{ubt}
\end{equation} 
where ${\cal U}_{\beta,\mu}$ is given as
\begin{equation*}
{\cal U}_{\beta,\mu}=e^{{\cal B}^{\dagger}(\beta,\mu)-{\cal B}(\beta,\mu)},
\label{ubm}
\end{equation*}
with 
\be
{\cal B}^\dagger(\beta,\mu) =\int \Big [
\sum_{n=0}^{\infty} \int d\vec k_{\omit x} 
q_r^\prime (n,k_z)^\dagger \theta_-(k_z,n, \beta,\mu)
\underline q_r^{\prime} (n,k_z)^\dagger +
\tilde q_r^\prime (n,k_z) \theta_+(k_z,n,\beta,\mu)
\underline { \tilde q}_r^{\prime} (n,k_z)\Big ].
\label{bth}
\ee
In Eq.(\ref{bth}),
the underlined operators are the operators in the extended Hilbert space
 associated with thermal doubling in TFD method, and, 
 the ansatz functions $\theta_{\pm}(n,k_z,\beta,\mu)$
are related to quark and antiquark distributions as can be seen through
the minimization of the thermodynamic potential. 
 In Eq.(\ref{bth}) we have suppressed
the color and flavor indices in the quark and antiquark operators 
 as well as in the functions
$\theta_\mp$.

 All the functions in the ansatz in Eq.(\ref{ubt})
are  obtained by minimizing the
thermodynamic potential.
We shall carry out this minimization
in the next section. However, before carrying out the minimization
procedure, let us  
focus our attention to the expectation values of some known operators 
to show that with the above variational ansatz for the `ground state' given in
Eq.(\ref{ubt}) these reduce to the already known expressions in the appropriate
limits.
 
 Let us first consider the expectation value of the chiral order
parameter. The 
expectation value for chiral order parameter for the $i$-th flavor is given as

\bearr
\langle\Omega(\beta,\mu)|\bar\psi_i\psi_i|\Omega(\beta,\mu)\rangle
 &=& -\sum_{n=0}^\infty\frac{N_c|q_i|B\alpha_n}{(2\pi)^2}
\int{\frac{dp_z}{\epsilon_{ni}}\left[m_i\cos{2f_i}+|\zbf p_i|\sin{2f_i}\right]}
\left(1-\sin^2\theta_-^i-\sin^2\theta_+^i\right)\nonumber\\
&\equiv&-I^i
\label{isi}
\eearr
where, $\alpha_n=(2-\delta_{n,0})$ is the degeneracy factor of the
 $n$-th Landau level (all levels are doubly degenerate except the lowest Landau level).
As we shall see later, the functions $\sin^2\theta_\mp$ will be related to the
distribution functions for the quarks and antiquarks. Further, for later convenience,
it is useful to define $\cos\phi_0^i=m_i/\epsilon_{ni}$ and $\sin\phi_0^i=|\zbf p_i|/
\epsilon_{ni}$ so as to rewrite the order parameter $I_i$ as
\be
I_i=
 \sum_{n=0}^\infty\frac{N_c|q_i|B\alpha_n}{(2\pi)^2}
\int{dp_z}\cos\phi^i \left(1-\sin^2\theta_-^i-\sin^2\theta_+^i\right),
\label{Ii}
\ee
where we have defined $\phi^i=\phi_0^i-2f_i$. As we shall see later, it is convenient to
vary $\phi^i$ as compared to the original condensate function $f_i$ given through 
Eq.(\ref{ansatz}). This expression for the order parameter, $I_i$,
in the limit of vanishing condensates ($f_i$=0)
reduces to the expression derived in Ref. \cite{kausik}. Further, the expression
for the chiral  condensate
in the absence of the magnetic field at zero temperature and zero density becomes
\be
\langle\bar\psi^i\psi^i\rangle=-I_i=-\frac{6}{(2\pi)^3}\int d\zbf p \cos\phi^i,
\label{nofield}
\ee
once one realizes that in presence of quantizing magnetic field with discrete
Landau levels, one has \cite{digal}
\begin{equation*}
\int \frac{d\zbf p}{(2\pi) ^3}\rightarrow 
\frac{|qB|}{(2\pi)^2}\sum_{n=0}^\infty\alpha_n\int dp_z.
\end{equation*}
This expression for the condensate, Eq.(\ref{nofield}) is exactly the same 
as derived earlier in the absence of the magnetic field \cite{hmspmnjl,hmam}.

The other quantity that we wish to investigate is the axial fermion current 
density that is induced at finite chemical potential including the effect of 
temperature. The expectation value of the 
axial current density is given by
\begin{equation*}
\langle j_5^3\rangle \equiv \langle\bar{\psi_i^a}\gamma^3\gamma^5\psi_j^a\rangle.
\end{equation*}
Using the field operator expansion Eq.(\ref{psip}) and Eq.(\ref{primed+vespinors})
for the explicit forms for the spinors, we have for the $i$-th flavor
\be
\langle j_5^{i3}\rangle=
 \sum_n\frac{N_c}{(2\pi)^2} \int{dp_{\omit x}\left(I_n^2-I_{n-1}^2\right)\left(\sin^2
\theta_-^i-\sin^2\theta_+^i\right)}.
\label{j35}
\ee

Integrating over $dp_y$ using the orthonormal condition of Eq.(\ref{orthoI}), 
all the terms in the above sum for the Landau levels cancel out except for the
 zeroth Landau level so that,
\be
\langle j_5^{i3}\rangle=
\frac{N_c|q_i|B}{(2\pi)^2}\int{dp_z\left[\sin^2\theta_-^{i0}-\sin^2\theta_+^{i0}\right]}.
\label{j35p}
\end{equation}
which is identical to that in Ref.\cite{metlitsky} once we identify the functions
$\sin^2\theta_\mp^{i0}$ as the particle and the 
 antiparticle distribution functions for the zero modes (see e.g.
Eq.(\ref{them}) in the next section).

\section{Evaluation of thermodynamic potential and gap equations }
\label{evaluation}
As has already been mentioned, we shall consider in the present investigation,
the 3-flavor 
Nambu Jona Lasinio model including the  Kobayashi-Maskawa-t-Hooft (KMT)
determinant interaction. The corresponding
Hamiltonian density is given as
\bearr
{\cal H} & = &\psi^ \dagger(-i\bfm \alpha \cdot \bfm \nabla-qBx\alpha_2
+\gamma^0 \hat m )\psi
-G_s\sum_{A=0}^8\left[(\bar\psi\lambda^A\psi)^2-
(\bar\psi\gamma^5\lambda^A\psi)^2\right]\nonumber\\
&+&K\left[{ det_f[\bar\psi(1+\gamma_5)\psi]
+det_f[\bar\psi(1-\gamma_5)\psi]}\right]
\label{ham}
\eearr
where $\psi ^{i,a}$ denotes a quark field with color `$a$' 
$(a=r,g,b)$, and flavor `$i$'
 $(i=u,d,s)$, indices. The matrix of current quark masses is given by
$\hat m$=diag$_f(m_u,m_d,m_s)$ in the flavor space.
We  shall assume  in the present investigation, isospin
symmetry with $m_u$=$m_d$.  
Strictly speaking, when the electromagnetic effects are taken into account,
the current quark masses of u and d quarks should not be the same due to the
difference in their electrical charges. However, because of the smallness of the
electromagnetic coupling, we shall ignore this tiny effect and continue with
$m_u=m_d$ in the present investigation of chiral symmetry breaking.
In Eq. (\ref{ham}), $\lambda^A$, $A=1,\cdots 8$ denote the Gellman matrices
acting in the flavor space and
$\lambda^0 = \sqrt{\frac{2}{3}}\,1\hspace{-1.5mm}1_f$,
$1\hspace{-1.5mm}1_f$ as the unit matrix in the flavor space.
The four point interaction term $\sim G_s$ is symmetric in $SU(3)_V\times
SU(3)_A\times U(1)_V\times U(1)_A$. In contrast, the determinant term
$\sim K$ which for the case of three flavors generates a six point
interaction  breaks $U(1)_A$ symmetry. In the absence of the magnetic field and
the mass term, the overall symmetry in the flavor space 
is $SU(3)_V\times SU(3)_A \times U(1)_V$. This
spontaneously breaks to $SU(3)_V \times U(1)_V$ implying
the conservation of the baryon number and the flavor number. The current
quark mass term introduces additional explicit breaking of chiral symmetry
leading to partial conservation of  the axial  current. Due to the
presence of magnetic field on the other hand  the $SU(3)_V$ symmetry
 in the flavor space reduces to to $SU(2)_V\times SU(2)_A$ since the u quark
 has different  electric charge compared
 to d and s quarks \cite{manfer}.

Next, we evaluate the expectation value of the kinetic term in Eq.(\ref{ham}) which is
given as

\be
T=\langle\Omega(\beta,\mu)|
\psi_i^{a \dag}(-i\vec\alpha\cdot\zbf\nabla-q_iBx\alpha_2)\psi_i^a|\Omega(\beta,\mu)
\rangle. 
\label{T}
\ee
To evaluate this we use Eq. (\ref{psip}) and the results of spatial
derivatives on the functions $I_n(\xi)$ ($\xi=\sqrt{|q_i|B}(x-{p_y}/(|q_i|B)))$. 
\begin{equation*}
\frac{\partial I_n}{\partial x} = \sqrt{|q_i|B}\left[-\xi I_n 
+ \sqrt{2n}I_{n-1}\right], 
\end{equation*}
\be
\frac{\partial I_{n-1}}{\partial x} = 
\sqrt{|q_i|B}\left[-\xi I_{n-1} + \sqrt{2(n-1)}I_{n-2}\right].
\ee
Using above, a straightforward but tedious manipulation leads to the
expression for the kinetic term as
\be
T
= -\sum_{n=0}^\infty\sum_i\frac{N_c\alpha_n|q_iB|}{(2\pi)^2}\int{dp_z(m_i\cos\phi_i + |\zbf p_i|\sin\phi_i)
(1 - \sin^2\theta_-^i - \sin^2\theta_+^i)}
\label{tt}
\ee

The contribution from the quartic interaction  term in Eq. (\ref{ham}), 
using Eq. (\ref{isi}) turns out to be, 
\begin{equation}
{V_S}\equiv -G_s \langle \Omega(\beta,\mu)|
\sum_{A=0}^8\left[(\bar\psi\lambda^A\psi)^2-
(\bar\psi\gamma^5\lambda^A\psi)^2\right]
| \Omega(\beta,\mu)\rangle
=-2 G_S\sum_{i=1,3}{I^i}^2,
\label{vs}
\end{equation}
where we have used the properties of the Gellman matrices $
\sum_{A=0}^8\lambda_{ij}^A\lambda_{kl}^A = 2\delta_{il}\delta_{jk}$.

Finally, the contribution from the six quark interaction term leads to the
energy expectation value as
\be
V_{det} =
+K\langle{ det_f[\bar\psi(1+\gamma_5)\psi]
+det_f[\bar\psi(1-\gamma_5)\psi]}\rangle\nonumber\\
=-2 K I_1I_2I_3.
\label{vdet}
\ee
The thermodynamic potential is then given by
\be
\Omega=T+V_S+V_{det}-\sum_{i=1}^3\mu_i\rho_i -\frac{1}{\beta}s
\label{Omega}
\ee
In the above, $\mu_{i} $ is the chemical potential for the quark of flavor $i$.
The total number density of the quarks is given by
\be
\rho=\sum_{i=1}^3\rho_i=\sum_i \langle\psi_i^\dag\psi_i\rangle
 = \sum_{n=0}^\infty\sum_i\frac{N_c\alpha_n|q_iB|}{(2\pi)^2}
\int{dp_z\left[\sin^2\theta_-^i - \sin^2\theta_+^i\right]}.
\label{rho}
\ee
Finally, for the entropy density for the quarks we have \cite{tfd}
\begin{equation}
s = -\sum_i\sum_n\frac{N_c\alpha_n|q_i|B}{(2\pi)^2}\int{dp_z\lbrace
(\sin^2\theta_-^i\ln{\sin^2\theta_-^i} +
\cos^2\theta_-^i\ln{\cos^2\theta_-^i}) + (-\rightarrow +)\rbrace}.
\label{entropy}
\end{equation}

Now the functional  minimization of  the thermodynamic potential
 $\Omega$ with respect to  the chiral condensate function 
$f _i (p_z)$ leads to
\be
\cot\phi_i = \frac{m_i + 4GI_i + 2K|\epsilon_{ijk}|I_j I_k}{|\zbf p_i|} = 
\frac{M_i}{|\zbf p_i|}.
\label{tan2h}
\ee
We have defined, in the above, the constituent quark mass $M_i$ for the $i$-th flavor as
\begin{equation}
M_i = m_i + 4GI_i + 2K|\epsilon_{ijk}|I_j I_k
\label{gapeq}
\end{equation}
Finally, the minimization of the thermodynamic potential with respect to the
thermal functions $\theta_{\pm}(\zbf k)$ gives
\be
\sin^2\theta_\pm^{i,n}=\frac{1}{\exp(\beta(\omega_{i,n}\pm\mu_i))+1},
\label{them}
\ee
where, $\omega_{i,n}=\sqrt{M_i^2+p_z^2+2n|q_i|B)}$ is the excitation energy with
the constituent quark mass $M_i$.

Substituting the solution for the
condensate function of Eq. (\ref{tan2h}) and the thermal function given in
Eq.(\ref{them}) back in Eq. (\ref{Ii}) yields the chiral condensate as
\be
-\langle\bar\psi_i\psi_i\rangle\equiv
I_i=
 \sum_{n=0}^\infty\frac{N_c|q_i|B\alpha_n}{(2\pi)^2}
\int{dp_z}\Big (\frac{M_i}{\omega_i}\Big) \left(1-\sin^2\theta_-^i-\sin^2\theta_+^i\right).
\label{Iis}
\ee
Thus Eq.(\ref{gapeq}) and Eq.(\ref{Iis}) define the self consistent mass 
gap equation for the $i$-th quark flavor. Using the solutions for the condensate function
as well as the gap equation Eq.(\ref{gapeq}), the thermodynamic potential given
in Eq.(\ref{Omega}) reduces to
\begin{eqnarray}
\Omega &=& -\sum_{n,i}\frac{N_c\alpha_n|q_iB|}{(2\pi)^2}\int{dp_z\omega_{i}} \nonumber\\
 &-& 
\sum_{n,i}\frac{N_c\alpha_n|q_iB|}{(2\pi)^2\beta}\int{dp_z[\ln{\lbrace 1+e^{-\beta(\omega_i-\mu_i)}\rbrace} +
 \ln{\lbrace 1+e^{-\beta(\omega_i+\mu_i)}\rbrace}]}\nonumber\\
&+&
 2G\sum_i I_i^2 + 4KI_1 I_2 I_3.\nonumber\\
\label{thermtmuB}
\end{eqnarray}

The zero temperature and the zero density contribution  of the thermodynamic potential
 ($\Omega(T=0,\mu=0)$)
in the above is ultraviolet divergent, which  is also transmitted to the
gap equation Eq.(\ref{gapeq}) through the integral $I_i$ in Eq. (\ref{Iis}).
In the zero field case ($B$=0) such integrals are regularized either 
by a sharp cutoff (a step function in $|\zbf p|$)
that is common in many effective theories like NJL model
\cite{klevansky,bubrep,hatkun} although one can also use a smooth regulator
\cite{wilczek,berges,noornah}. The choice of the regulator is a part of the definition 
of the model with the constraint that 
the physically meaningful results should not eventually
be dependent on the regularization prescription. A sharp cutoff in presence of
the magnetic field suffers from cutoff artifact since the continuous
momentum dependence in two spatial dimensions are being replaced by a sum
over discretized Landau levels. To avoid this, a smooth parametrization
was used in Ref. \cite{fukushimaplb} in the context of chiral magnetic effects
 in Polyakov loop NJL model. In the present work  however we follow the elegant
procedure that was followed in Ref. \cite{providencia} by adding and subtracting
a vacuum (zero field) contribution to the  thermodynamic potential which is also
divergent. This manipulation makes the first term of Eq.(\ref{thermtmuB}) acquire a
physically more appealing form by separating the vacuum contribution and the 
finite field contribution written in terms of Riemann-Hurwitz $\zeta$ functions as
\bearr
 &&-\sum_{i=1}^{3}\sum_{n=0}^{\infty}
\frac{N_c\alpha_n|q_iB|}{(2\pi)^2}\int{dp_z}\sqrt{p_z^2+2n|q_i|B+M_i^2}\nonumber\\
&=& -\frac{2N_c}{(2\pi)^3}\sum_{i=1}^3\int{d\zbf p\sqrt{\zbf p^2 + M_i^2}}\nonumber\\
 &-& \frac{N_c}{2\pi^2}\sum_{i=1}^3|q_iB|^2\left[\zeta^\prime(-1,x_i) 
- \frac{1}{2}(x_i^2-x_i)\ln{x_i} + \frac{x_i^2}{4}\right],
\label{t1}
\eearr
where, we have defined the dimensionless quantity,
$x_i = \frac{M_i^2}{2|q_iB|}$, i.e. the mass parameter in units of the magnetic field.
Further, $\zeta^\prime(-1,x)=d\zeta(z,x)/dz|_{z=1} $ is the derivative of the 
Riemann-Hurwitz zeta function \cite{hurwitz}.

Using Eq.(\ref{t1}), the quark antiquark condensate of Eq.(\ref{Iis}) 
can also be separated
into a zero field (divergent) vacuum term, a (finite) field dependent term and
 a (finite) medium dependent term as
\bearr
-\langle\bar\psi_i\psi_i\rangle\equiv I_i&=&\frac{2N_c}{(2\pi)^3}\int 
{d\zbf p}\frac{M_i}{\sqrt{\zbf p^2+M_i^2}}\nonumber\\
 &+& \frac{N_c M_i|q_iB|}{(2\pi)^2}\left[x_i(1-\ln{x_i}) + \ln{\Gamma(x_i)} +
\frac{1}{2}\ln{\frac{x_i}{2\pi}}\right]\nonumber\\
& -&\sum_{n=0}^\infty\frac{N_c|q_i|B\alpha_n}{(2\pi)^2}
\int dp_z\frac{M_i}{\sqrt{M_i^2+|\zbf p_i|^2}} (\sin^2\theta_-^i +\sin^2\theta_+^i)\nonumber\\
 &=& I_{vac}^i + I_{field}^i+I_{med}^i,
\label{Iif}
\eearr
where, we have denoted the three terms in above equation as $I_{vac}$, $I_{field}$
and $I_{med}$ respectively.
The zero field vacuum contributions in Eq.(\ref{t1}) as well as in the Eq.(\ref{Iif}),
can be calculated with a sharp  three momentum cut off as is usually done in the NJL model
\cite{klevansky,bubrep}. Thus, e.g.,
 the vacuum part of the order parameter $I_{vac}$ becomes
\be
I^i_{vac}=\frac{N_cM_i}{2\pi^2}\left[\Lambda \sqrt{\Lambda^2+M_i^2}-M_i^2\log\left(
\frac{\Lambda+\sqrt{\Lambda^2+M_i^2}}{M_i}\right)\right].
\label{Ivac}
\ee
However, since in presence of magnetic field, $|\zbf p|^2=
p_z^2+2 n |q_iB|$, the condition of sharp three momentum cutoff translates
 to a finite number of Landau level summations in Eq.(\ref{Iif}) or in Eq.(\ref{t1}) 
with $n_{max}$, the maximum number of Landau levels that are filled up being given
as $n_{max}={\rm Int}\left[\frac{\Lambda^2}{2|q_i|B}\right]$ when the component of the
 momentum in the $z$-direction $p_z=0$.
Further, for the medium contribution $I_{med}$, this 
also leads to
a cut off for the magnitude of $|p_z|$ as $\Lambda^\prime=\sqrt{\Lambda^2-2n|q_i|B}$
for a given value of $n$.

Thus the thermodynamic potential,  given by Eq.(\ref{thermtmuB}),
can be rewritten as
\be
\Omega(\beta,\mu,B,M_i)=\Omega_{vac}+\Omega_{field}+\Omega_{med}+2G\sum_{i=1}^3I_i^2
+4 KI_1I_2I_3,
\label{thpot1}
\ee

where,
\be
\Omega_{vac}=-2N_c\sum_i\int_{|\zbf p|<\Lambda}
\frac{d\zbf p}{(2\pi)^3}\sqrt{\zbf p^2+M_i^2}
\equiv -\frac{N_c }{8\pi^2}\sum_i\left[(\Lambda^2+ M_i^2)^{1/2}(2\Lambda^2+ M_i^2)-
M_i^4\log\frac{\Lambda+\sqrt{\Lambda^2+M_i^2}}{ M_i}\right].
\ee

The field contribution to thermodynamic potential is given by
\be
\Omega_{field}=
 - \frac{N_c}{2\pi^2}\sum_{i=1}^3|q_iB|^2\left[\zeta^\prime(-1,x_i) 
- \frac{1}{2}(x_i^2-x_i)\ln{x_i} + \frac{x_i^2}{4}\right].
\label{omegafield}
\ee

The derivative of the Riemann-Hurwitz zeta function $\zeta(z,x)$ at
$z=-1$ is given by \cite{hurwitz}
\be
\zeta^\prime(-1,x) =-\frac{1}{2} x\log x-\frac{1}{4} x^2+\frac{1}{2} x^2\log x+
\frac{1}{12}\log x+x^2 \int_0^\infty \frac{2\tan^{-1} y+y \log(1+y^2)}{\exp(2\pi x y)-1} dy.
\ee

The medium contribution to the thermodynamic potential is
\be
\Omega_{med}=
\sum_{n,i}\frac{N_c\alpha_n|q_iB|}{(2\pi)^2\beta}\int{dp_z[\ln{\lbrace 1+e^{-\beta(\omega_i-\mu_i)}\rbrace} +
 \ln{\lbrace 1+e^{-\beta(\omega_i+\mu_i)}\rbrace}]}.
\ee

It may be useful to write down the zero temperature limits of the integrals
$\Omega_{med}$ and $I_{med}$. Let us note that at zero temperature the 
particle distribution function $\sin^2\theta_-=\Theta(\mu_i-\omega_{in})$ while 
the antiparticle distribution function $\sin^2\theta_+=0$. The $\theta$-function 
restricts the magnitude of $|p_z|$ to be less than
 $p_{zmax}^i=\sqrt{p_f^{i2}-2n|q_i|B}$, where, $p_f^i=\sqrt{\mu_i^2-M_i^2}$ is
the Fermi momentum of the corresponding flavor.
Further, this also restricts maximum number of Landau levels $n_{\max}$ to
$n_{max}^{if}={\rm Int}[\frac{p_f^{i2}}{2|q_i|B}]$. The contribution
arising due to the medium to the chiral condensate then reduces to
\be
I^i_{med}(T=0,B,\mu_i,M_i)=
\frac{N_c}{2\pi^2}\sum_{n=0}^{n_{max}^{if}}\alpha_n|q_i|BM_i
\log\left(\frac{p_{zmax}^i+\mu_i}{\sqrt{M_i^2+2n|q_i|B}}\right).
\label{Imedt0}
\ee
Similarly, the contribution from the medium to the thermodynamic potential at zero temperature
reduces to
\be
\Omega_{med}(T=0,B,\mu_i,M_i)
=\sum_i
\frac{N_c}{4\pi^2}\sum_{n=0}^{n_{max}^{if}}\alpha_n|q_i|BM_i
\left[\mu_ip_{zmax}^i-(M_i^2+2n|q_i|B)
\log\left(\frac{p_{zmax}^i+\mu_i}{\sqrt{M_i^2+2n|q_i|B}}\right)\right].
\label{omegamedt0}
\ee

In the context of neutron star matter,
the quark phase that could be present in the interior,
consists of the u,d,s quarks as well as electrons, 
in weak equilibrium 
\begin{subequations}
\be
d\rightarrow u+e^-+\bar\nu_{e^-},
\end{equation}
\be
s\rightarrow u+e^-+\bar\nu_{e^-},
\ee
and,
\begin{equation}
s+u \rightarrow d+u,
\end{equation}
\end{subequations}
leading to the relations between the chemical potentials 
$\mu_u$,$\mu_d$,$\mu_s$,$\mu_E$
as
\be
\mu_s=\mu_d=\mu_u+\mu_E.
\ee
 The neutrino chemical potentials are taken to be zero as they can diffuse out
of the star. So there are {\em two} independent chemical potentials needed
to describe the matter in the neutron star interior which we take to be the
quark chemical potential $\mu_q$
(one third of the baryon  chemical potential) and the electric charge chemical
potential, $\mu_e$ in terms of which the chemical potentials are given by
$\mu_s=\mu_q-\frac{1}{3}\mu_e =\mu_d$, $\mu_u=\mu_q+\frac{2}{3}\mu_e$ and
$\mu_E=-\mu_e$. In addition, for description of the charge neutral matter, there is a
further constraint for the chemical potentials through the following relation for the
particle densities given by
\be
\frac{2}{3}\rho_u-\frac{1}{3}\rho_d-\frac{1}{3}\rho_s-\rho_E=0.
\label{neutrality}
\ee
The quark number densities $\rho_i$ for each flavor are already defined in Eq.(\ref{rho})
and the electron number density is given by
\be
\rho_E = \sum_{n=0}^\infty\frac{1\alpha_n|eB|}{(2\pi)^2}
\int{dp_z\left[\sin^2\theta_-^e - \sin^2\theta_+^e\right]},
\label{rhoe}
\ee
where, the distribution functions for the electron, 
$\sin^2\theta_\mp =1/(\exp(\omega_e\mp\mu_E)+1)$, with $\omega_e=\sqrt{p_z^2+2n|e|B}$.

To calculate the total thermodynamic potential (negative of the pressure)
relevant for neutron star one has to add the thermodynamic potential
$\Omega_e$ due to the electrons, to the thermodynamic potential 
for the quarks as given in Eq.(\ref{thermtmuB}). The contribution of the electrons is given by
\be
\Omega_e=
 \sum_{n,i}\frac{\alpha_n|eB|}{(2\pi)^2\beta}\int{dp_z[\ln{\lbrace 1+
e^{-\beta(\omega_e-\mu_E)}\rbrace} +
 \ln{\lbrace 1+e^{-\beta(\omega_i+\mu_E)}\rbrace}]}
\label{omge}
\ee
The thermodynamic potential (Eq. (\ref{thpot1})), the mass gap equations Eq. (\ref{gapeq}),
Eq. (\ref{t1}) and the charge neutrality condition, Eq.(\ref{neutrality}) are the
basis for our numerical calculations for various physical situations that we shall
 discuss in the following section.

\section{results and discussions}

For numerical calculations, we have taken the values of
the parameters of the NJL model as follows. The coupling constant
$G_s$ has the dimension of $[{\rm Mass}]^{-2}$ while 
the six fermion coupling $K$
has a dimension $[{\rm Mass}]^{-5}$. To regularize the divergent integrals
we use a sharp cut-off, $\Lambda$ in 3-momentum space. Thus we have 
five parameters in total, namely the current quark masses
for the non strange and strange quarks, $m_q$ and $m_s$,
the two couplings $G_s$, $K$ and the three-momentum cutoff $\Lambda$.
We have chosen here $\Lambda=0.6023$ GeV, $G_s\Lambda^2=1.835$,
$K\Lambda^5=12.36$, $m_q=5.5$ MeV and $m_s=0.1407$ GeV  as has been 
used in Ref.\cite{rehberg}. After choosing $m_q=5.5$ MeV, the remaining four
parameters are fixed by fitting to the pion decay constant and
the masses of pion, kaon and $\eta'$. With this set of parameters the
mass of $\eta$ is underestimated by about six percent and 
the constituent masses of the
light quarks turn out to be $M_1=0.368$ GeV 
for u-d quarks, while the same for strange quark turns out as $M_s=0.549$ GeV,
at zero temperature and zero density. 
It might be relevant here to
comment regarding the choice of the parameters.
There have been different sets of
parameters by other groups also \cite{hatkun,lkw,bubrep} for
the three flavor NJL model. Although the same
principle  as above is used, e.g., as in Ref \cite{hatkun}, the resulting
parameter sets are not identical. In particular, the dimensionless
coupling $K\Lambda^5$ differs by as large as about 30 percent as compared to
the value used here. This discrepancy is due to a different treatment
of the $\eta' $ meson. Since NJL model does not confine, and because
of the large mass of the $\eta'$ meson ($m_{\eta'}$=958 MeV), it lies
above the threshold for $q\bar q $ decay with an unphysical
imaginary part for the corresponding polarization diagram. 
This is an unavoidable feature of NJL model and leaves an uncertainty
which is reflected in the difference in the parameter sets by 
different groups. Within this limitation regarding the parameters 
of the model, however, we proceed with the above parameter set 
which has  already been used in the study of the phase diagram of 
dense matter in Ref. \cite{ruester} as well as in the context of 
equation of state for neutron star matter in Ref. \cite{leupold}.

\begin{figure}
\vspace{-0.4cm}
\begin{center}
\begin{tabular}{c c }
\includegraphics[width=8cm,height=8cm]{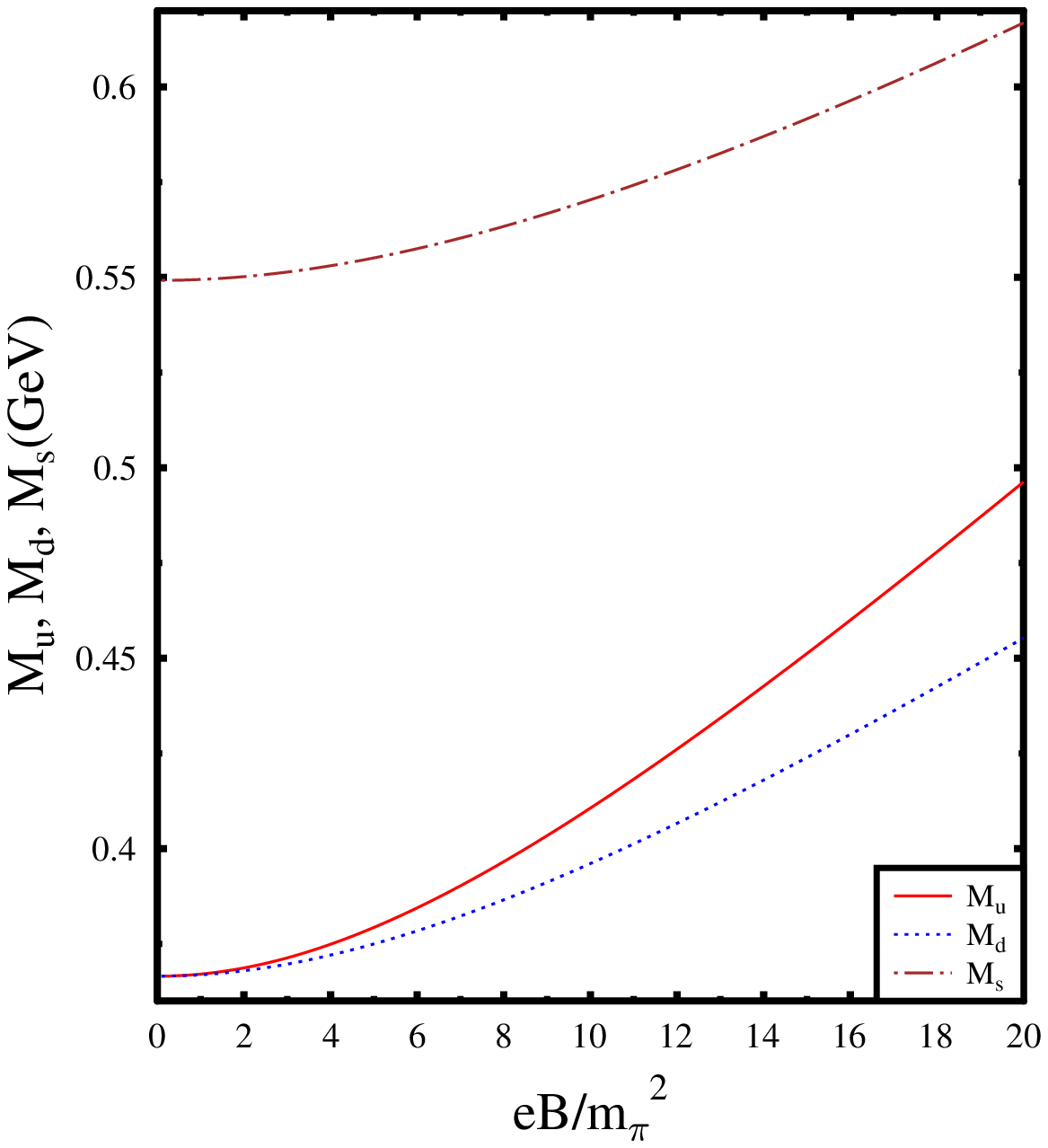}&
\includegraphics[width=8cm,height=8cm]{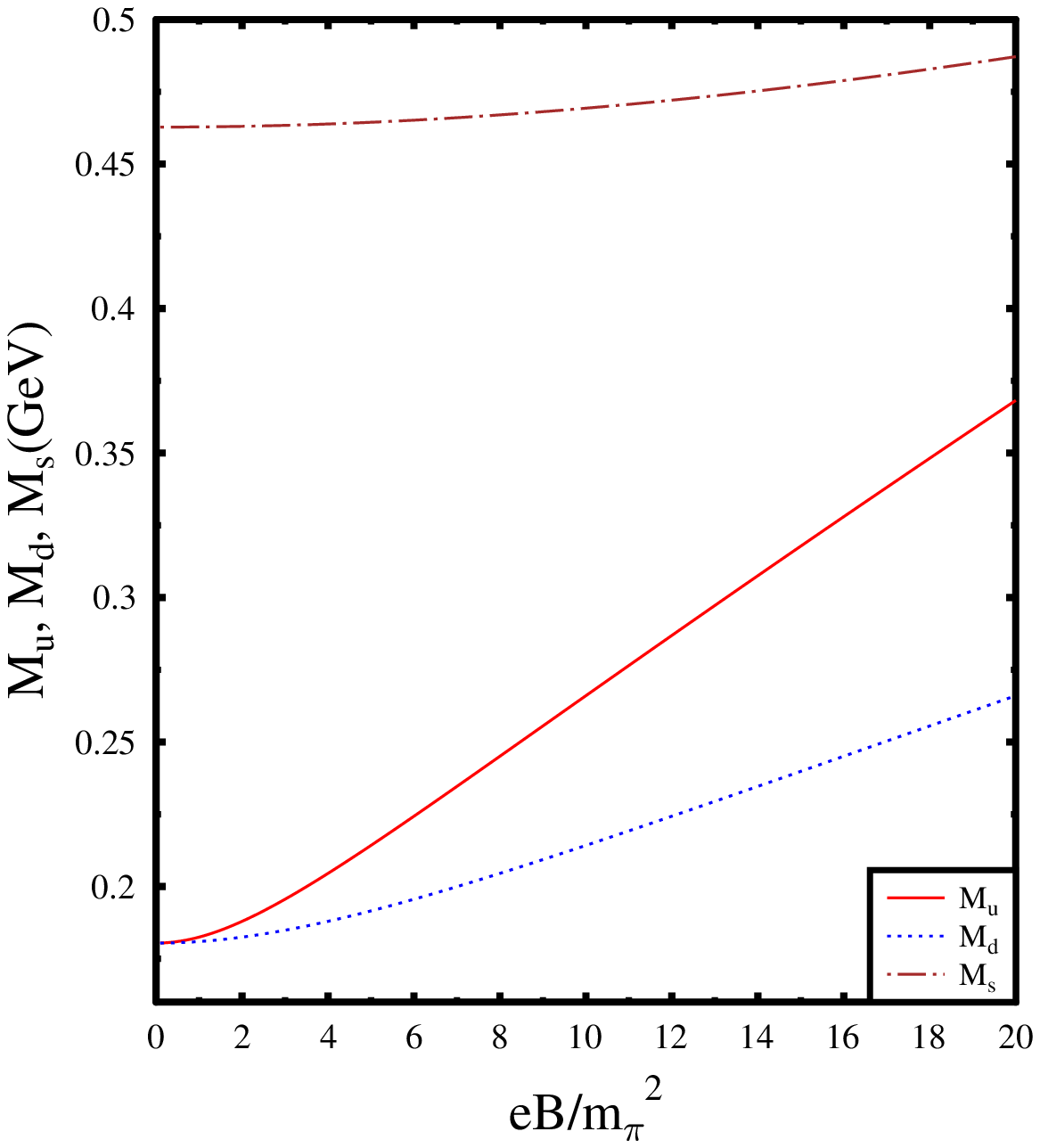}\\
Fig. 1-a & Fig.1-b
\end{tabular}
\end{center}
\caption{Constituent quark masses as functions of magnetic field at
zero temperature and zero density. Fig. 1(a) shows the constituent quark masses
as a function of the magnetic field when the determinant interaction
is taken into account. Fig 1(b) shows the same when the determinant interaction term
is ignored i.e. $K=0$. The solid curves refer to constituent masses of u- quark, the
dotted curve refers to the constituent mass of the d-quarks while the
dot-dashed curve refers to the same of the strange quark.
}
\label{fig1}
\end{figure}

\begin{figure}
\vspace{-0.4cm}
\begin{center}
\begin{tabular}{c c }
\includegraphics[width=8cm,height=8cm]{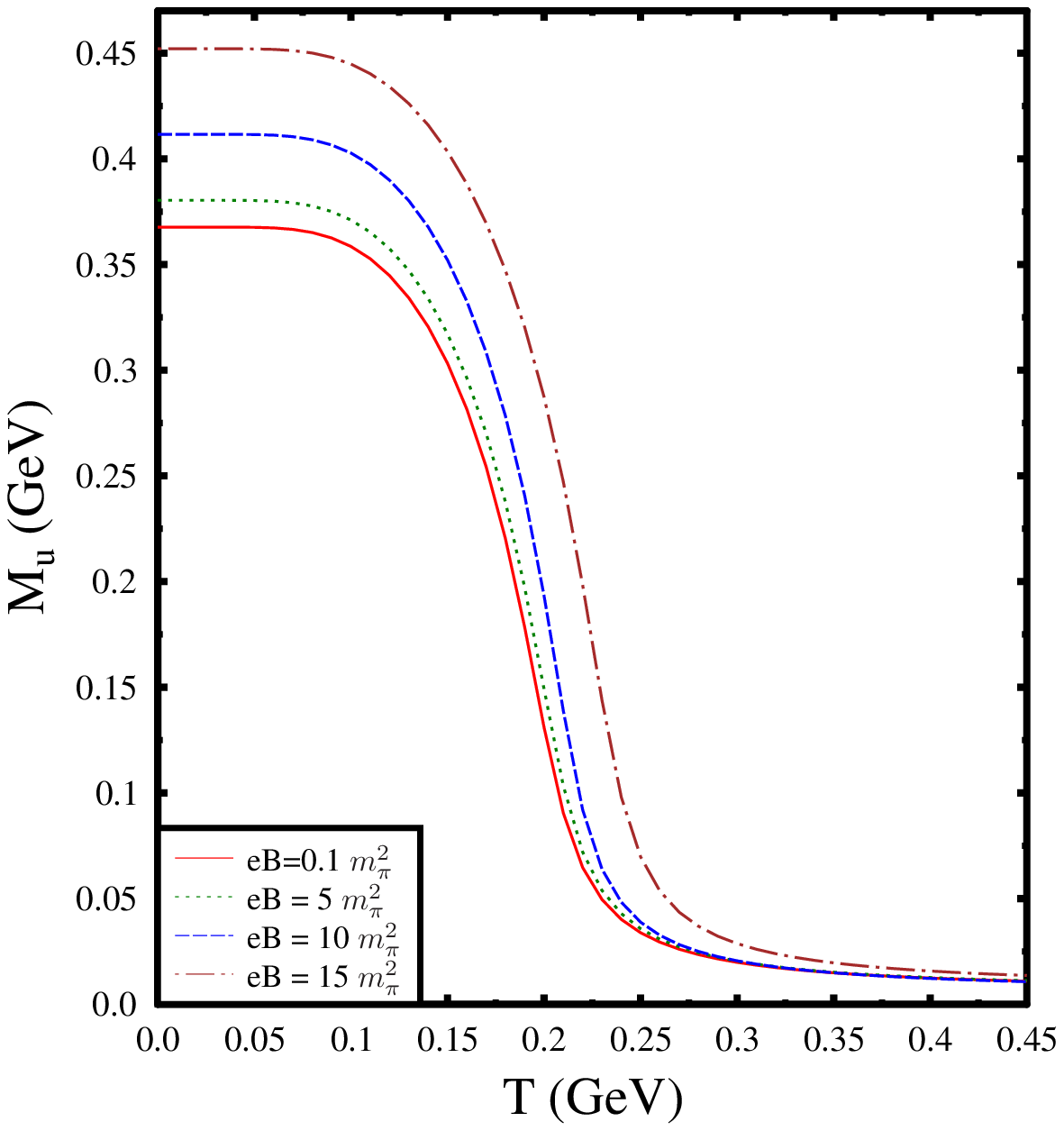}&
\includegraphics[width=8cm,height=8cm]{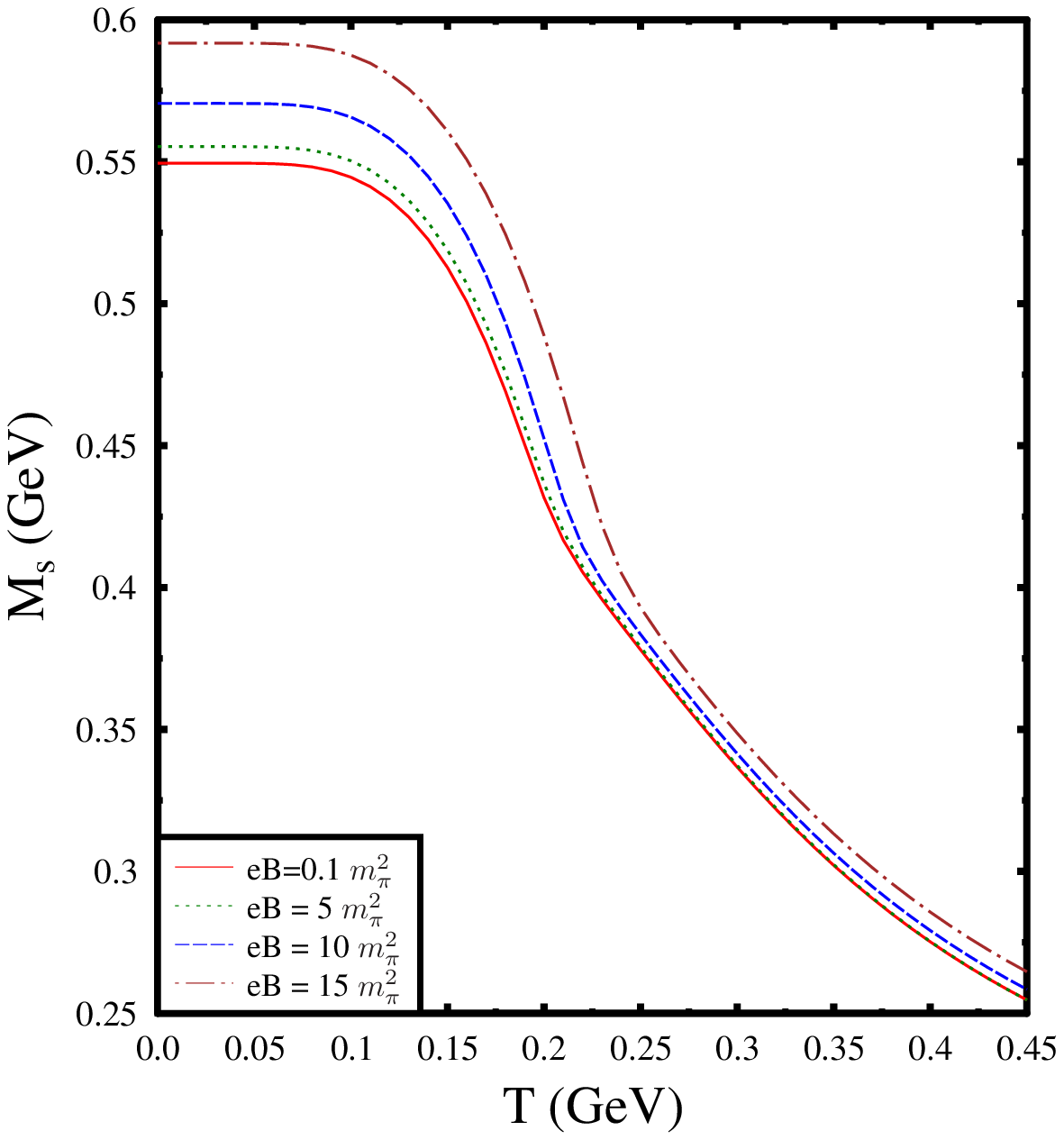}\\
Fig. 2-a & Fig.2-b
\end{tabular}
\end{center}
\caption{Constituent quark masses when charge neutrality 
conditions are not imposed.
Fig.2-a shows the $M_u$  at zero baryon chemical potential
 as a function of temperature for different values of the magnetic field.
Fig. 2-b  shows the same for the strange quark mass $M_s$.
Both the subplots correspond to nonzero values for the current quark
masses given as $m_u$=5.5 MeV and $m_s$=140.7 MeV.
}
\label{fig2}
\end{figure}

Let us begin the discussion of the results for the case when the charge
neutrality condition is not imposed. In this  case $\mu_e=0$ and
all the quark flavors have the same chemical potential $\mu_q$. For given values of
$\mu_q$, $T(=\beta^{-1})$ and $eB$, we solve the
mass gap equation Eq.(\ref{gapeq}) self consistently using the expression Eq.(\ref{Iif})
for the order parameter.
Few comments regarding the evaluation of $I^i_{med}$ of Eq.(\ref{Iif})
may be worth mentioning. In these evaluations,
while considering zero temperature and nonzero $\mu_q$, the Landau levels are filled
up upto a maximum value of $n$,  $n_{max}={\rm Int}
\left[\frac{\Lambda^2}{2|q_i|B}\right]$ 
as already mentioned in the previous section. On the other hand,
for all finite temperature
calculations, the levels are filled  upto  the maximum Landau level, so that the error
in neglecting the higher Landau level is less than $10^{-5}$. 
 We also observe that for low temperatures,  near the cross over 
transition temperature, there could be 
multiple solutions of the mass gap equation
corresponding to multiple extrema of the thermodynamic potential.
In such cases, we have chosen the solution which has the least value of the
 thermodynamic potential given in Eq.(\ref{thpot1}). We ensure this 
by verifying the positivity
of the second derivative of the thermodynamic potential with respect
to the corresponding masses.

We  show the constituent masses of the three flavors of quarks as modified
by magnetic field at zero temperature and zero density in Fig.\ref{fig1}. 
The magnetic field enhances the order parameters as
reflected in the values of the constituent masses $M_u$, $M_d$ and $M_s$. Because of
charge difference, this enhancement is not the same for all the quarks. For the couplings
$G$ and $K$ as chosen here, the enhancement factors $ (M(B)-M(B=0))/M(B=0)$ e.g.
 for $eB=20m_\pi^2$ are about 35$\%$, 24$\%$, 12$\%$ for u,d and s quarks 
respectively. We might mention here that the effect of magnetic field on 
chiral symmetry breaking has been considered
in NJL model in Ref.\cite{klimenkoplb} without the KMT determinant 
interaction term.  For a comparison, we have also plotted in Fig. 1 b, 
the constituent quark masses without the determinant term as a function of 
the magnetic field. Clearly, the mass splitting between $u$ and $d$ quarks
is much larger  when the determinant interaction is not taken
in to account-- e.g.$(M_u(B)-M_d(B))/M_u(B=0)=57\%$ at $eB=20m_\pi^2$ when K=0 while the
same ratio is about $11\%$ when $K\Lambda^5=12.36$. This behavior can be understood 
as follows. Whereas the magnetic field tends to differentiate the constituent
quark masses of different flavors, the determinant interaction which causes mixing
between the constituent quarks of different flavors tends to bring the constituent
quark masses together. This results in the splitting between the constituent quarks
of different flavors becoming smaller when determinant interaction is included in presence of magnetic field. Such a behavior is also observed in Ref.\cite{boomsma} for the case
of two flavor NJL model.

We then show the temperature dependence of the
constituent quark masses of the u and s quark for zero chemical
potential for different strengths of the magnetic field,
in Fig.\ref{fig2}. 
The phase transition
remains a smooth crossover as is the case with zero magnetic field.
The effect of magnetic field as a catalyser of chiral symmetry breaking is 
also evident. The chiral condensate and hence the constituent
quark masses increase in the temperature regime considered here
when the magnetic field is
increased. In these calculations all the Landau levels as appropriate for 
the given magnetic field have been filled up and the lowest Landau
level approximation has not been assumed. 
The qualitative aspects of the phase transition remains
the same as in case of zero magnetic field.
This result is in contrast to
the linear sigma model coupled to quarks \cite{fraga} where the usual cross over
becomes a first order phase transition in the presence of strong magnetic field.
We might mention here that our results are similar to that of Ref.\cite {boomsma}
 for the two flavor NJL model.

\begin{figure}
\vspace{-0.4cm}
\begin{center}
\begin{tabular}{c c c }
\includegraphics[width=6cm,height=6cm]{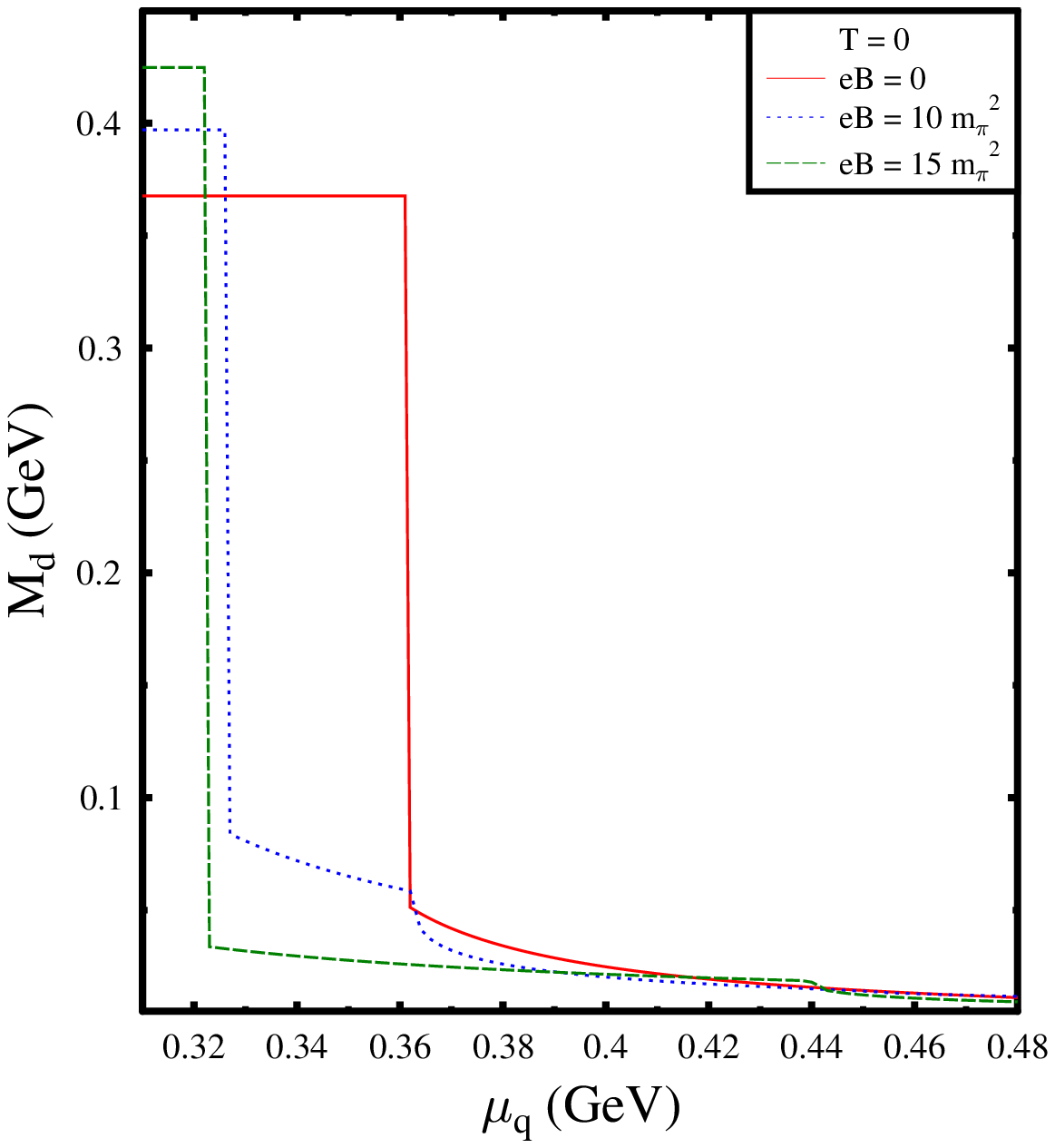}&
\includegraphics[width=6cm,height=6cm]{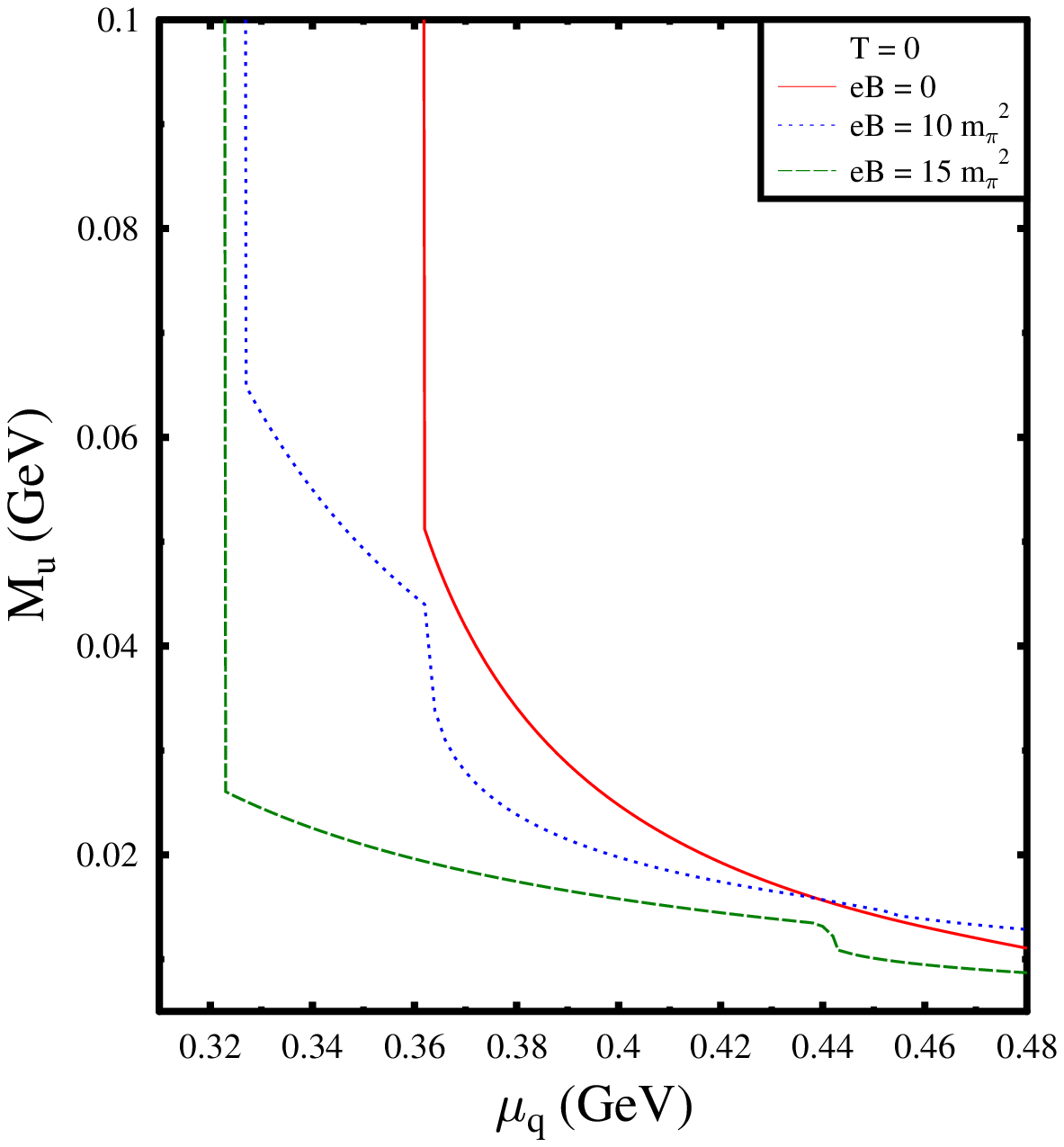} &
\includegraphics[width=6cm,height=6cm]{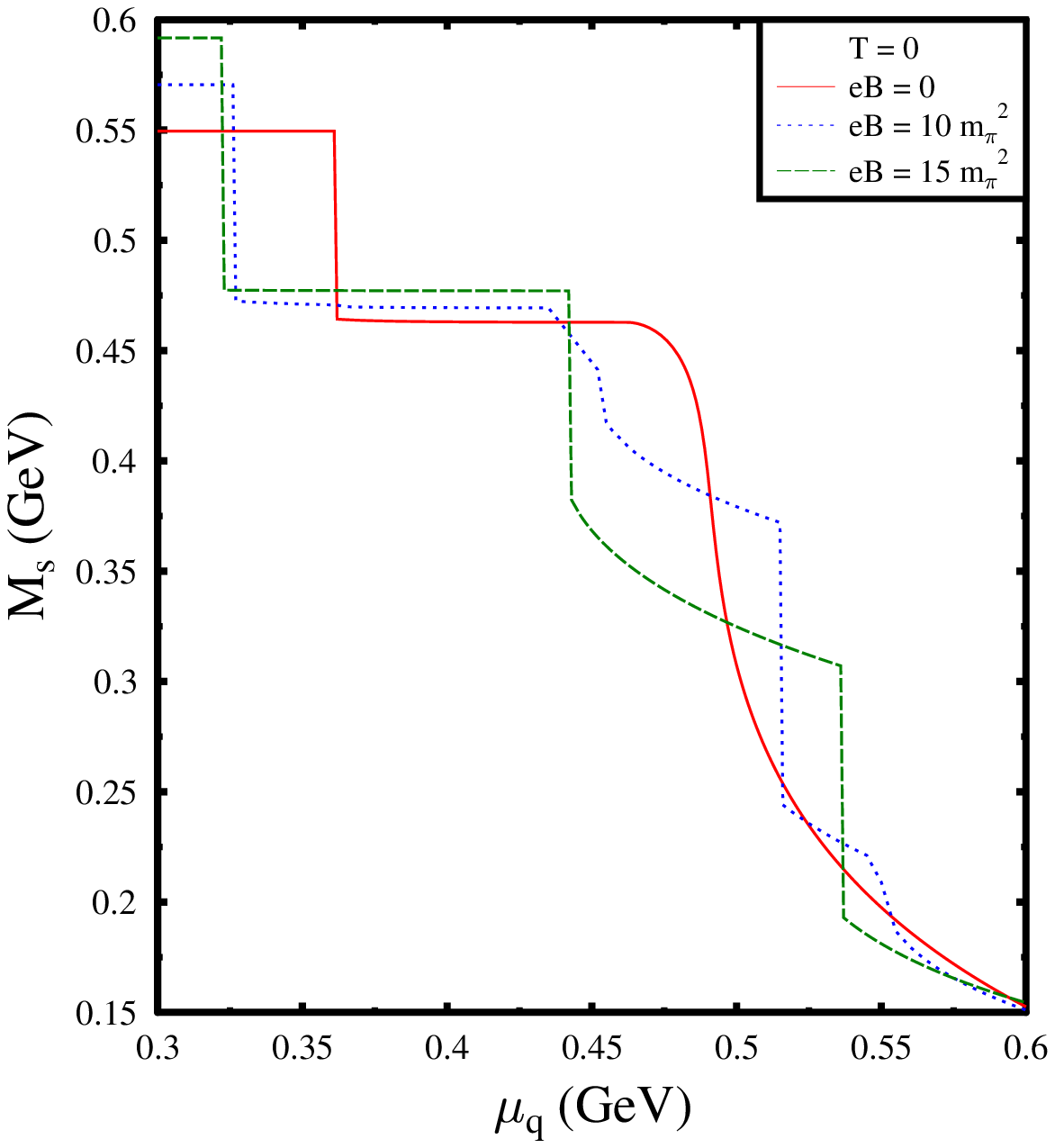}\\
Fig. 3-a & Fig.3-b & Fig.3-c
\end{tabular}
\end{center}
\caption{Constituent quark masses as functions of $\mu_q$ at  $T=0$ for different
strength of magnetic field. Constituent quark masses of d,u and s quarks are plotted
in Fig a, Fig b and Fig c respectively.
}
\label{fig3}
\end{figure}

Next, we discuss the behavior of constituent masses as baryon number
density is increased at zero temperature and for
different strengths of magnetic field. 
In Fig.\ref{fig3},
we show the dependence of the constituent masses $M_u$, $M_d$ and $M_s$
 on the quark chemical
potential at zero temperature for three different
strengths of magnetic field.
For zero magnetic field, as the quark chemical potential is increased,
a first order transition is observed to take place for the value
of $\mu_q=\mu_c\sim 0.362$GeV. At lower values of the quark chemical
 potential ($\mu_q<\mu_c)$,
the masses of the quarks stay at their vacuum values and the baryon 
number density remains zero.
At $\mu=\mu_c$, the first order transition takes place and the light
quarks have a drop in their masses from their vacuum values of about 367 MeV to
about 52 MeV. The baryon number
density also jumps from zero to $2.37 \rho_0$, with $\rho_0= 0.17{\rm fm}^{-3}$ 
being the normal nuclear matter density. Because of the six fermion KMT term,
this first order transition for the light quarks is also reflected in dropping 
of the strange quark mass from its vacuum value of 549 MeV to about 464 MeV. 

\begin{figure}
\includegraphics[width=8cm,height=8cm]{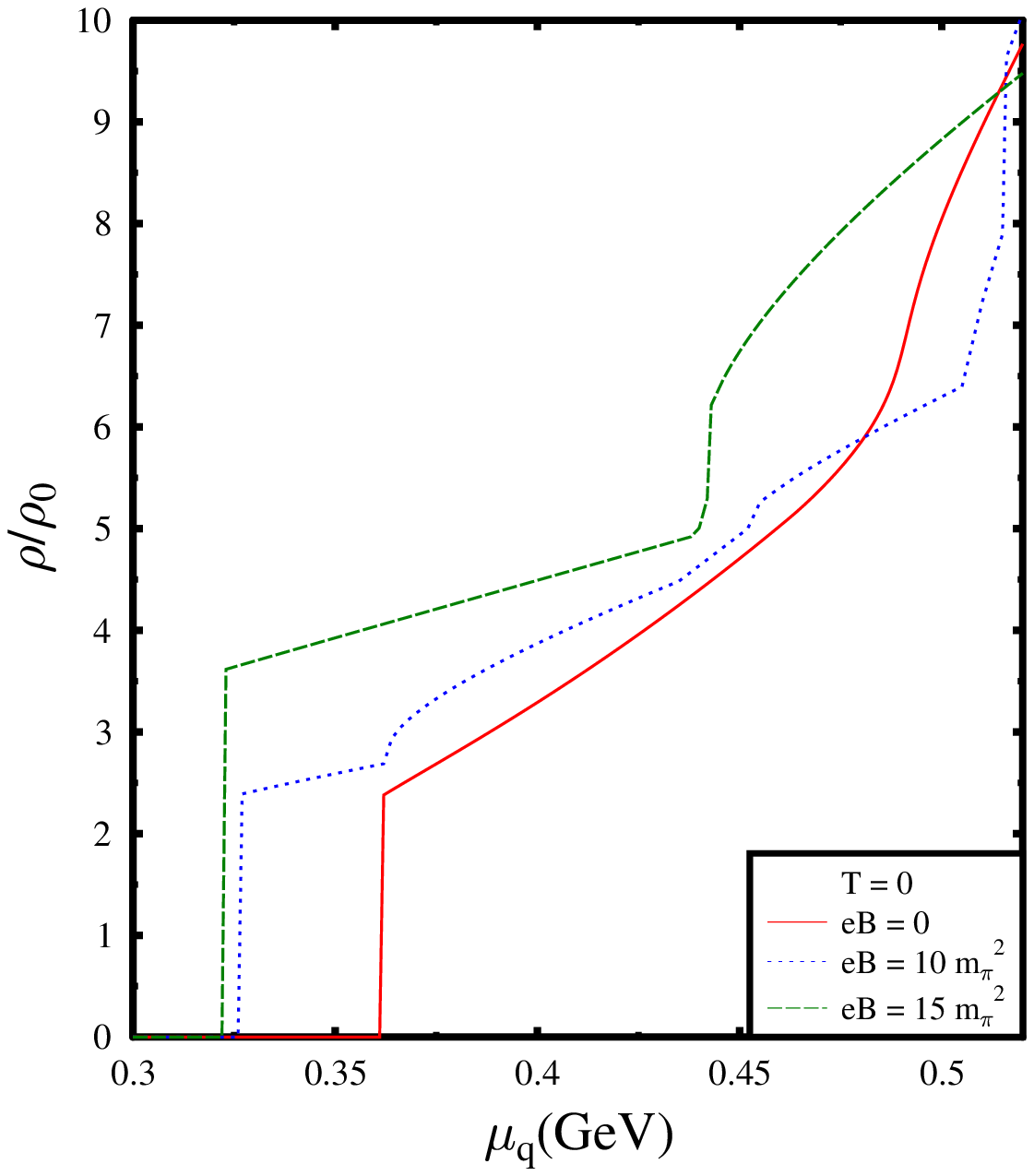}
\caption{ Baryon density as a function of quark chemical potential
for different strengths of magnetic field.
}
\label{fig4}
\end{figure}

As the magnetic field is increased, the critical chemical potential for this first
order transition
consistently decreases as may be clear from figures Fig. 3a, Fig 3b and Fig 3c.
For $eB=10m_\pi^2$,$eB=15m_\pi^2$, the corresponding values of $\mu_c$ are 0.327 GeV
and 0.323 GeV respectively.
For $\mu<\mu_c$, the constituent quark masses
increases with the magnetic field as may be seen in Fig. 3a where we have plotted
the d-quark mass as a function of quark chemical potential.
E.g. for $eB=10m_\pi^2$, the increase in masses  of the u and d quarks are about 
45 MeV and 30 MeV respectively while for strange quarks the corresponding 
increase in mass is about 21 MeV as compared to zero field case.  Since the $\mu_c$
decreases with increase in magnetic field, there are windows in the range
of chemical potential where it appears e.g. in Fig 3a for the u-quark, 
that the mass decreases with the magnetic field in the range of chemical
 potential between $\mu=323$ MeV to
$\mu=362$ MeV. In this regime however, the chiral transition already has taken 
place for $eB=15m_\pi^2$. For $eB=10m_\pi^2$, in this regime of
chemical potential although the first order transition takes place, the transition
is weaker compared to the case of $eB=15m_\pi^2$ in the sense that the 
constituent quark mass is 
higher compared to the case of $eB=15m_\pi^2$.
Finally, for the case of zero magnetic field, the transition is still
to take place. After the transition the ordering in the masses
changes depending upon the filling up of the Landau levels in case of
 nonzero magnetic fields.
 The kinks in the mass variation correspond to filling up
of the Landau levels. This decrease of critical chemical potential due to
the presence of magnetic field has also been observed in dense holographic 
matter and is termed as inverse magnetic catalysis of chiral symmetry 
breaking \cite{andreasrebhan}. We, however, observe that although the critical chemical
potential decreases with magnetic field, the corresponding baryonic density
increases. This is clearly seen in  Fig. \ref{fig4} where we have shown 
the baryon number density as a function of quark chemical potential for different 
strengths of magnetic field. While for $eB=10m_\pi^2$, the critical 
density $\rho_c/\rho_0=2.39$
is almost similar to the zero field value of the same $\rho_c/\rho_0=2.38$,
the critical density for $eB=15m_\pi^2$ is substantially larger 
with $\rho_c/\rho_0=3.62$.  Similar qualitative behavior was also 
observed in Ref.\cite{providencia}.

\begin{figure}
\includegraphics[width=8cm,height=8cm]{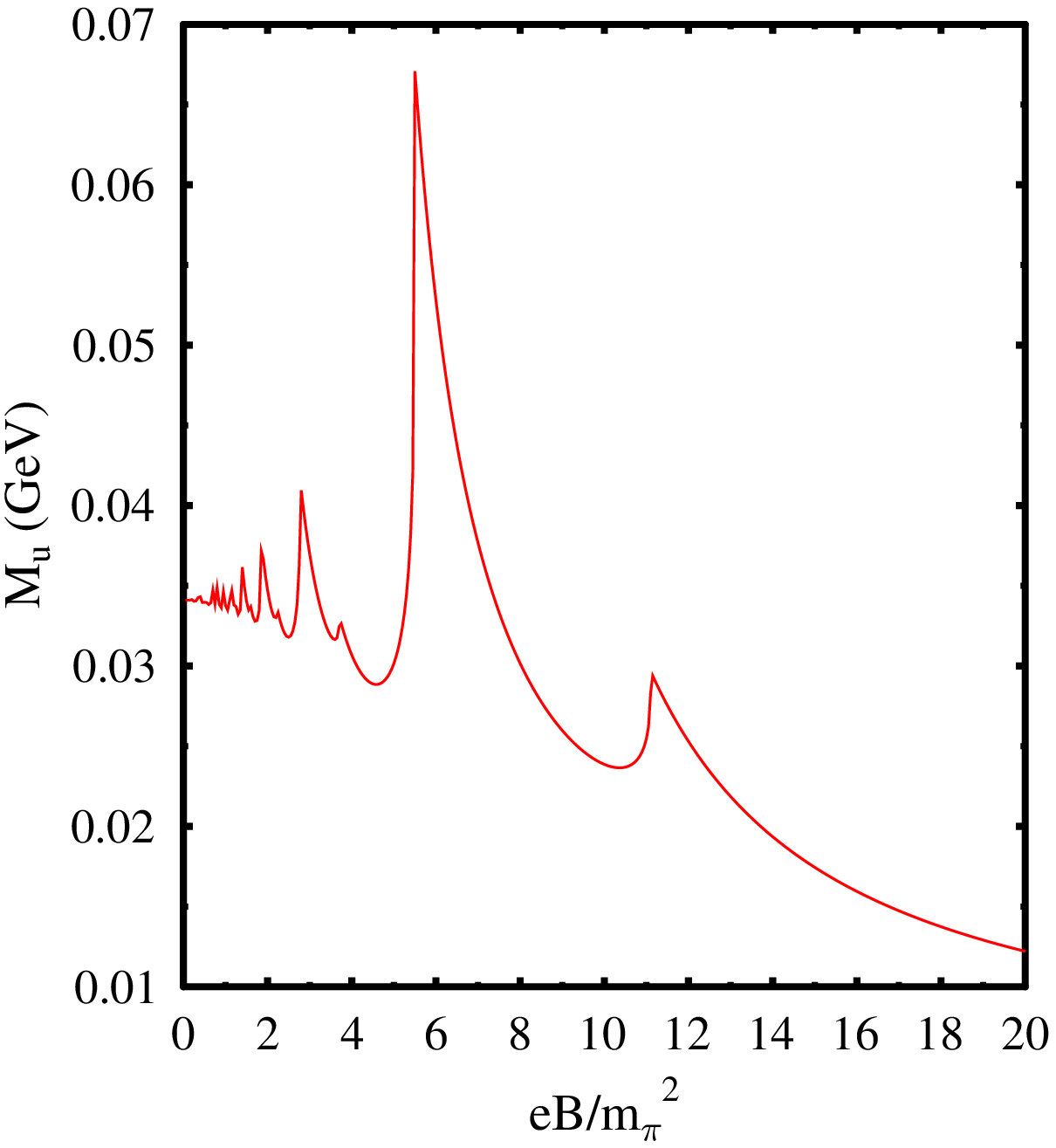}
\caption{ Oscillation of u-quark mass with magnetic field. We have
taken $\mu_q =380 $ MeV and T=0.}
\label{fig5}
\end{figure}

At finite chemical potential, we also observe oscillations of the order parameter
with the magnetic field as shown in Fig.\ref{fig5} for u-quark. 
We have taken the value of the
chemical potential as $\mu_q=380$ MeV and taken the temperature, T as zero.
This phenomenon is similar to the 
oscillation of the magnetization of a material in presence of external
magnetic field,  known as de Hass van Alphen effect \cite{ebertmag}. 
As observed earlier, this
is a consequence of oscillations in the density of states at Fermi surface due to the
Landau quantization. The oscillatory behavior is seen as long as 
$2 |q|B< \sqrt{\mu_q^2-M_q^2}$ and ceases when the first Landau level lies above the
Fermi surface \cite{fukuwarringa,noornah}.

\begin{figure}
\vspace{-0.4cm}
\begin{center}
\begin{tabular}{c c }
\includegraphics[width=8cm,height=8cm]{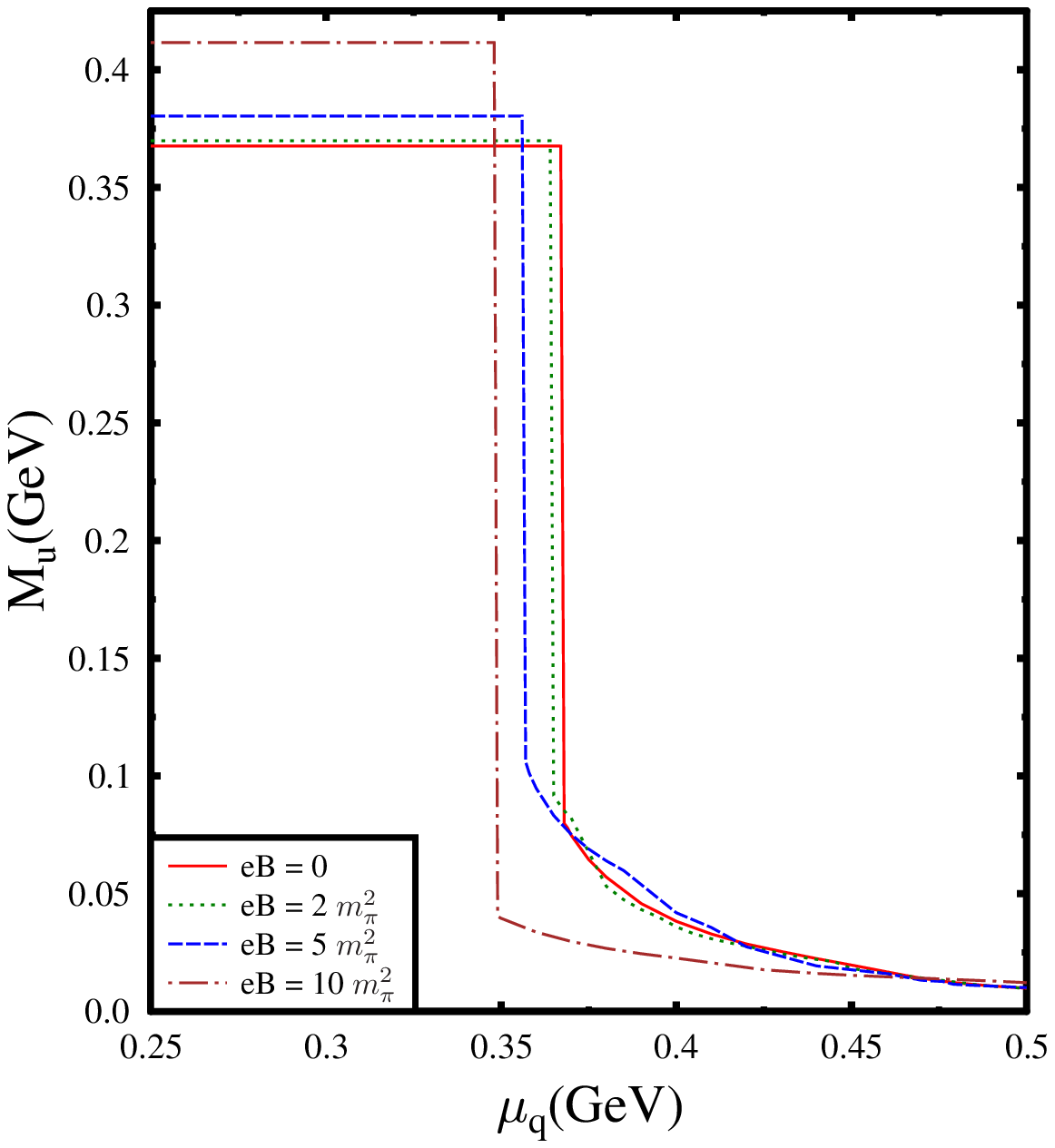}&
\includegraphics[width=8cm,height=8cm]{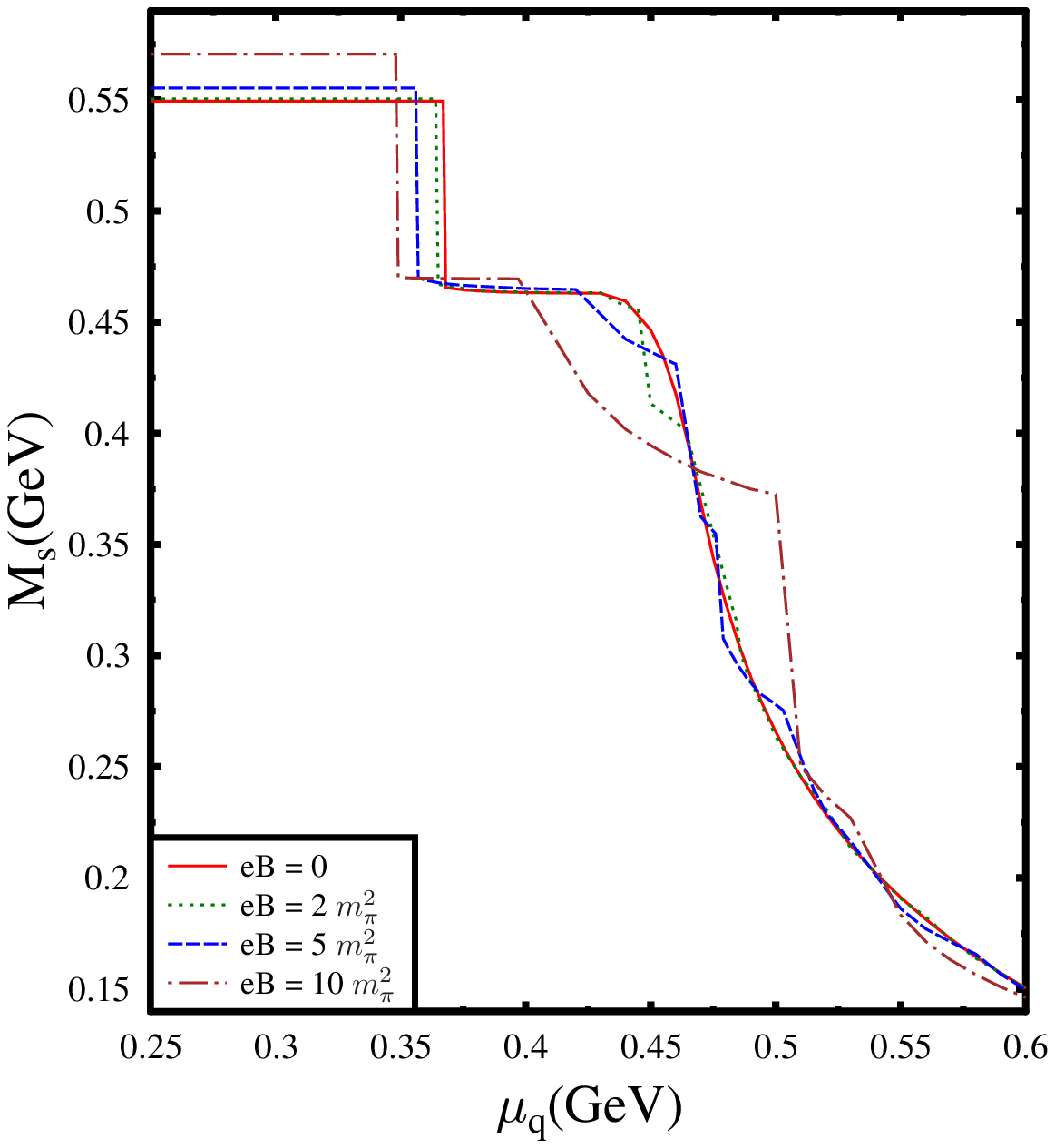}\\
Fig. 6-a & Fig.6-b
\end{tabular}
\end{center}
\caption{Constituent quark masses for charge neutral matter.
Masses of up quarks (a)  and strange quarks (b) as functions of the quark chemical
potential, $\mu_q$ at T=0 for different
strengths of magnetic field}
\label{fig6}
\end{figure}

\begin{figure}
\vspace{-0.4cm}
\begin{center}
\begin{tabular}{c c }
\includegraphics[width=8cm,height=8cm]{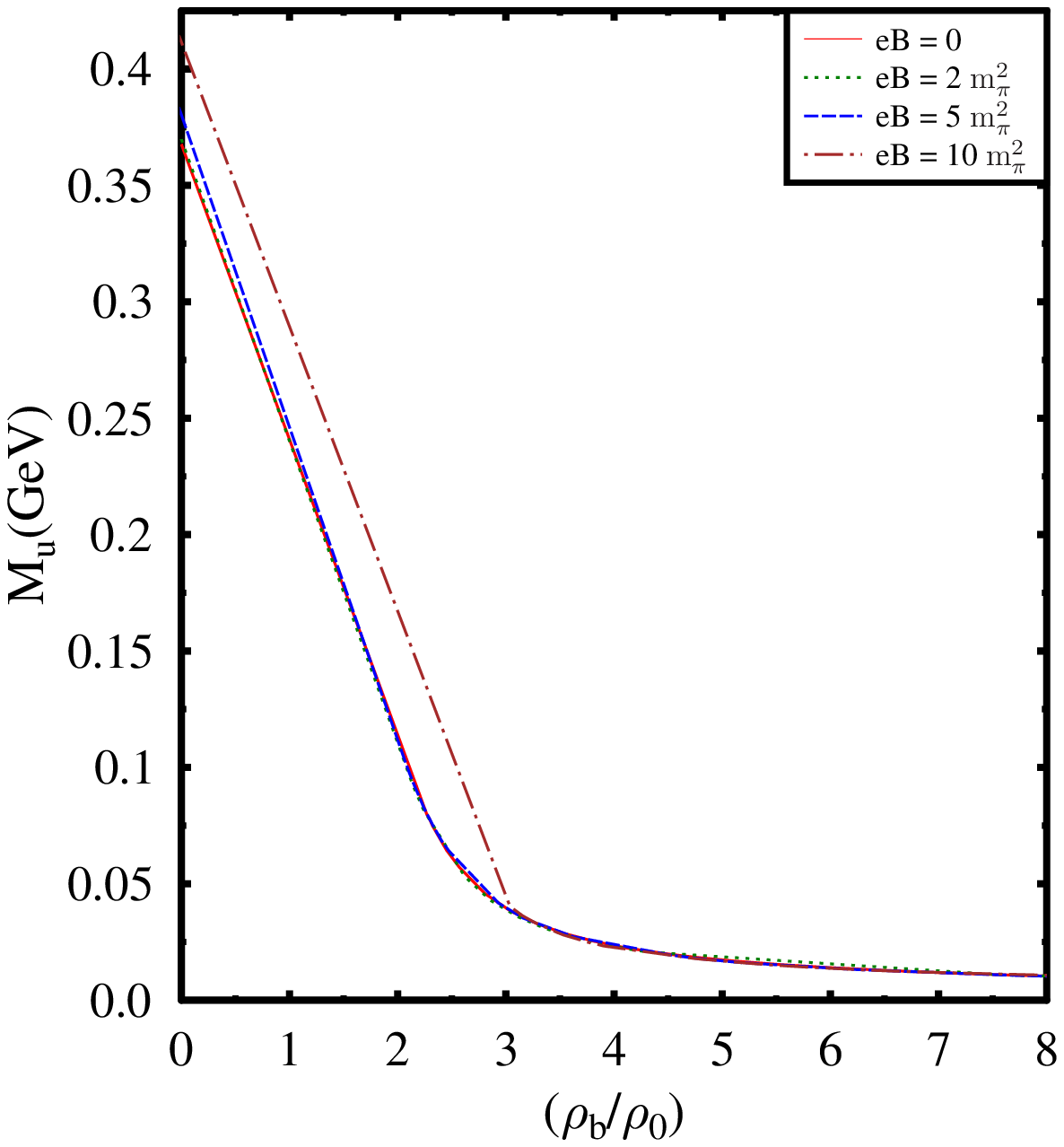}&
\includegraphics[width=8cm,height=8cm]{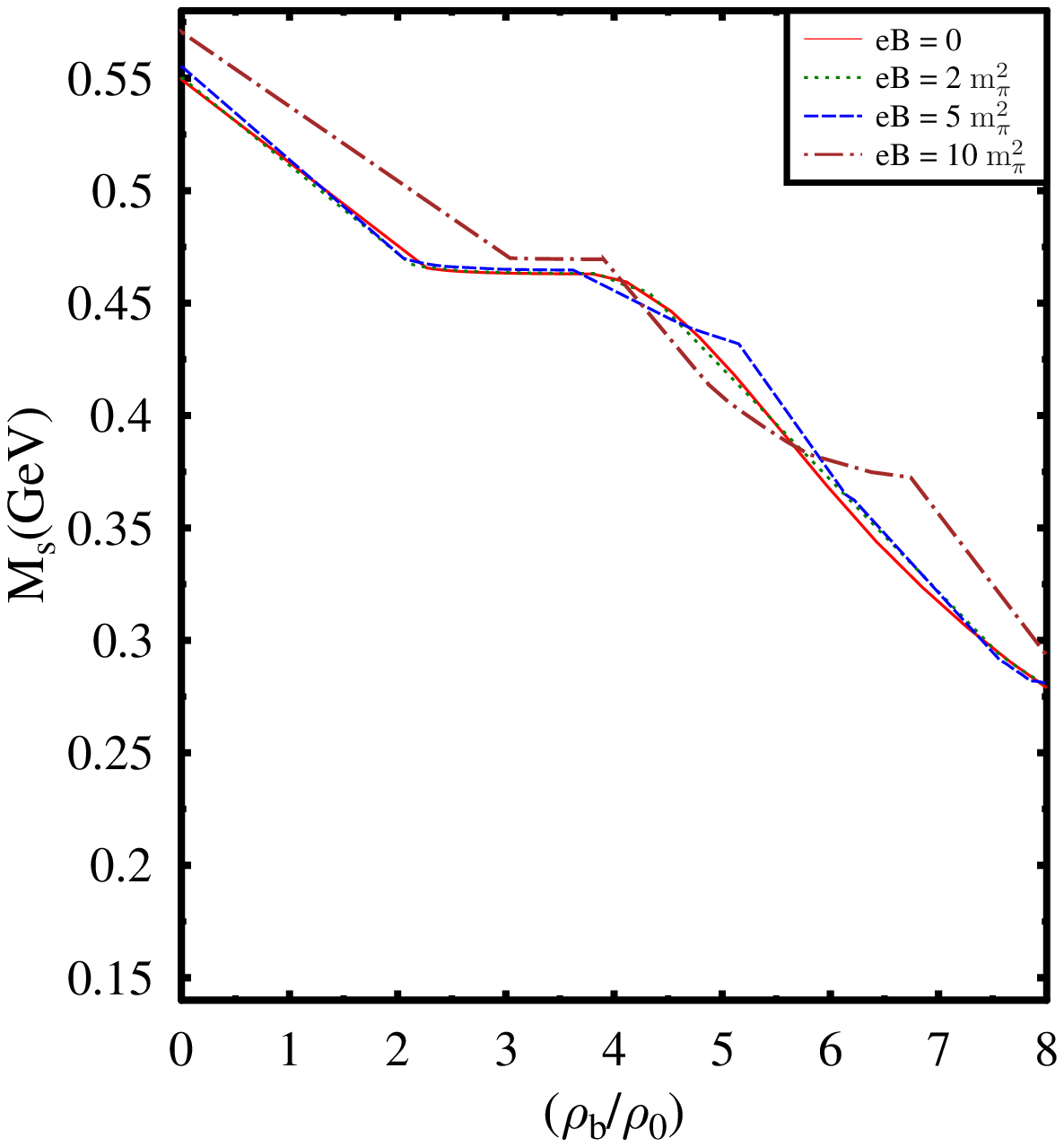}\\
Fig. 7-a & Fig.7-b
\end{tabular}
\end{center}
\caption{Constituent quark masses of
u and s quark masses as  functions of baryon density in units of
nuclear matter density $\rho_0$ for different
strengths of magnetic field at T=0.}
\label{fig7}
\end{figure}

We then discuss the effects of magnetic field on charge neutral dense matter 
as may be relevant for the matter in the interior of the neutron stars. The 
thermodynamic potential is numerically computed as follows. For  given values 
of the quark chemical potential $\mu_q$,
the electric charge chemical potential $\mu_E$ and magnetic field $eB$, the coupled 
mass gap equations given by Eq.(\ref{gapeq}) are solved, using the expression 
Eq.(\ref{Iif}) for the 
order parameter. The values of the 
electric charge chemical potential $\mu_E$ are varied so that the charge 
neutrality condition Eq. (\ref{neutrality}) is satisfied. The resulting solutions are
then used in Eq.(\ref{thpot1}) to compute the thermodynamic potential. In doing so,
we also check if there are multiple solutions to the gap equation and choose
the one that has the least value of the 
thermodynamic potential. In Fig. \ref{fig6}, we show the
masses of the quarks as functions of chemical potential for charge neutral matter for
zero temperature.
 Let us note that at the transition point, the d-quark
number density is almost twice that of u-quark number density to maintain charge
neutrality as mass of s-quark is much too large to contribute to the
charge density. For this to be realized, the mass of d-quark should be
sufficiently smaller as compared the mass of u quark to generate the
required difference in the number densities. This, in turn, means that
$\mu_d$ should be larger than $\mu_u$ unlike the charge neutral case
where all the quarks have the same chemical potential. Numerically,
it turns out that for zero magnetic field this condition is
satisfied when $\mu_d\sim 393$ MeV and $\mu_u\sim 318$ MeV as compared to
the common chemical potential $\mu_c=362$ MeV when charge neutrality 
condition is not imposed. The corresponding masses of d and u 
quarks are about $61 {\rm MeV}$
 and $80 {\rm MeV}$ respectively compared to
the common mass of $52 {\rm MeV} $ at the critical chemical potential
 when charge neutrality is not imposed. 
These values of $\mu_u$ and $\mu_d$ at the transition point corresponds to a
electron chemical potential of $\mu_e\sim 75$ MeV at the transition
while the corresponding quark chemical potential at the transition is 
$\mu_q=368 {\rm MeV} \equiv\mu_c$,
which is slightly higher as compared to the value
of $\mu_c=362 {\rm MeV}$ for the case when such neutrality condition is not
imposed. At the transition point for the neutral
matter, the number density of d quarks is almost twice that of
u quarks while the electron number density is three orders of magnitude lower than
either of the quark number densities.
As the magnetic field is increased, the constituent masses for the 
three quarks increase for chemical potential smaller than
the critical chemical potential. The first sudden drop
of the strange quark mass (see Fig.6b) is related to the drop in the
light quark masses through the determinant interaction.
The kink structure in the strange quark mass for higher magnetic fields
can be identifiable with the filling of different Landau levels.
Further, it is observed that a higher magnetic field leads to a 
smaller value for $\mu_c$. Similar to the case of non charge neutral matter,
however, the critical density becomes higher for increased magnetic field.
This magnetic catalysis of chiral symmetry breaking is clearly shown in
Fig.\ref{fig7}, where masses of up and strange quarks are shown as functions
of baryon density for zero temperature and for different strengths of magnetic field.

\begin{figure}
\includegraphics[width=8cm,height=8cm]{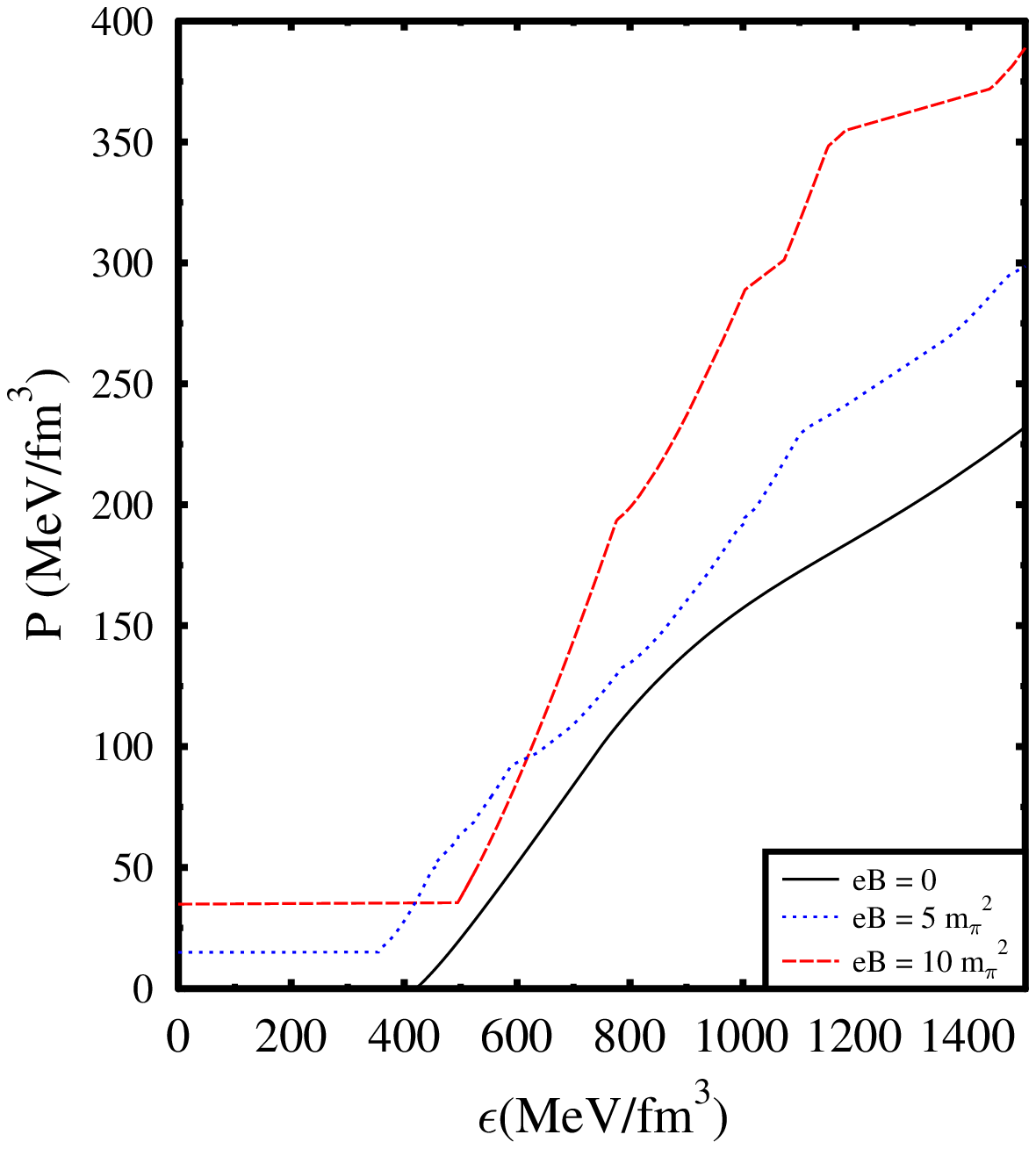}
\caption{Equations of state for charge neutral matter at zero temperature
for different strengths of the magnetic field.}
\label{fig8}
\end{figure}

We then study the effect of magnetic field on the equation of state, i.e.
pressure as a function of energy for the charge neutral matter. This is shown in Fig.
\ref{fig8} for zero temperature. The effect of Landau quantization shows up
in the kink structure of the equation of state. For smaller magnetic fields, this effect
is less visible as the number of filled Landau levels are quite large. Further,
it may be observed that as the magnetic field is increased, 
the equation of state
becomes somewhat stiffer. Since the zero density constituent quark masses increase with
magnetic field, the vacuum energy density becomes lower compared to the
zero field case. Therefore the starting values of pressure in presence of field
 becomes higher compared to the zero field case as seen in Fig.\ref{fig8}.  
For higher densities when chiral symmetry is restored, let us first note that
the contribution to the
thermodynamic potential due to the
magnetic field as given in Eq.(\ref{omegafield}) increases with magnetic field.
Therefore, it means that one has to
have a larger chemical potential for lower magnetic field as compared to higher field 
to have the same energy density. So one would naively expect the pressure 
($P=\mu\rho-\epsilon$)
with lower field to be higher. However, one has to note that chiral symmetry is restored 
at a lower chemical potential for higher magnetic field as may be clear from 
Fig.\ref{fig6} and hence the number densities can become higher leading
to a higher pressure. This is what 
actually happens for larger energy densities for the two
fields shown in Fig.\ref{fig8}. Because of the lower critical chemical potential for
the case of $eB=10m_\pi^2$, the masses of the quarks are smaller and hence 
the densities becomes higher compared to the case of $eB=5m_\pi^2$ for the 
same energy densities leading to a stiffer equation of state.

\begin{figure}
\vspace{-0.4cm}
\begin{center}
\begin{tabular}{c c }
\includegraphics[width=8cm,height=8cm]{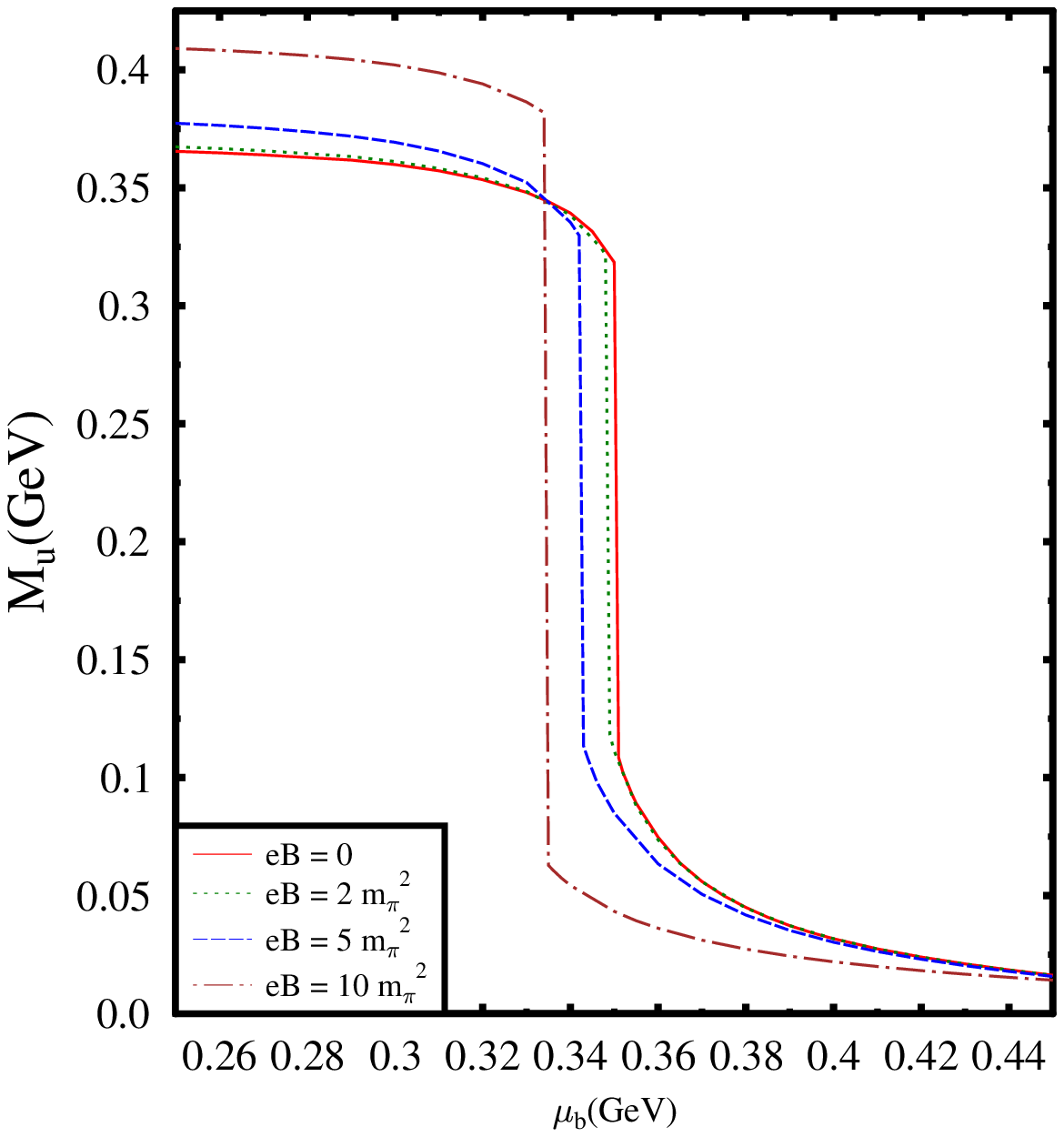}&
\includegraphics[width=8cm,height=8cm]{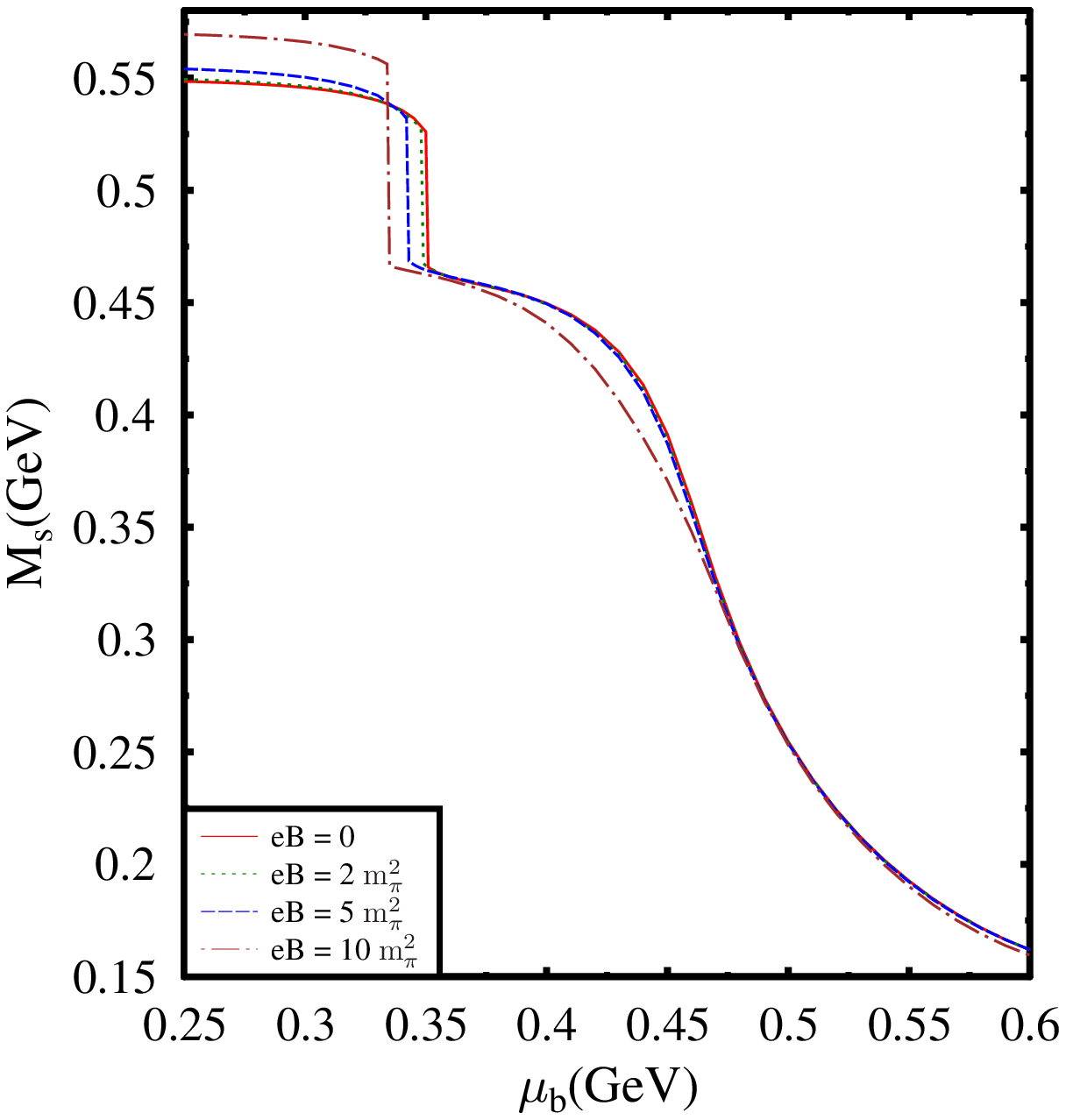}\\
Fig. 9-a & Fig.9-b
\end{tabular}
\end{center}
\caption{Constituent masses of u and s quarks as functions of $\mu_q$ for
charge neutral matter for different
strengths of magnetic field at T=40 MeV}
\label{fig9}
\end{figure}

Next, we discuss the effects of magnetic field on hot neutral quark matter. Such 
a condition is relevant for the matter in the interior of the
proto neutron stars where the temperatures 
could be of about few tens of MeV. In Fig.\ref{fig8}, we show the effect of temperature
on the masses of the quarks in the magnetized neutral matter. As may be expected,
the effect of temperature smoothes the behavior of the masses as functions 
of the quark chemical
potential. The corresponding equations of state are also shown 
in Fig.\ref{fig9}.
\begin{figure}
\includegraphics[width=8cm,height=8cm]{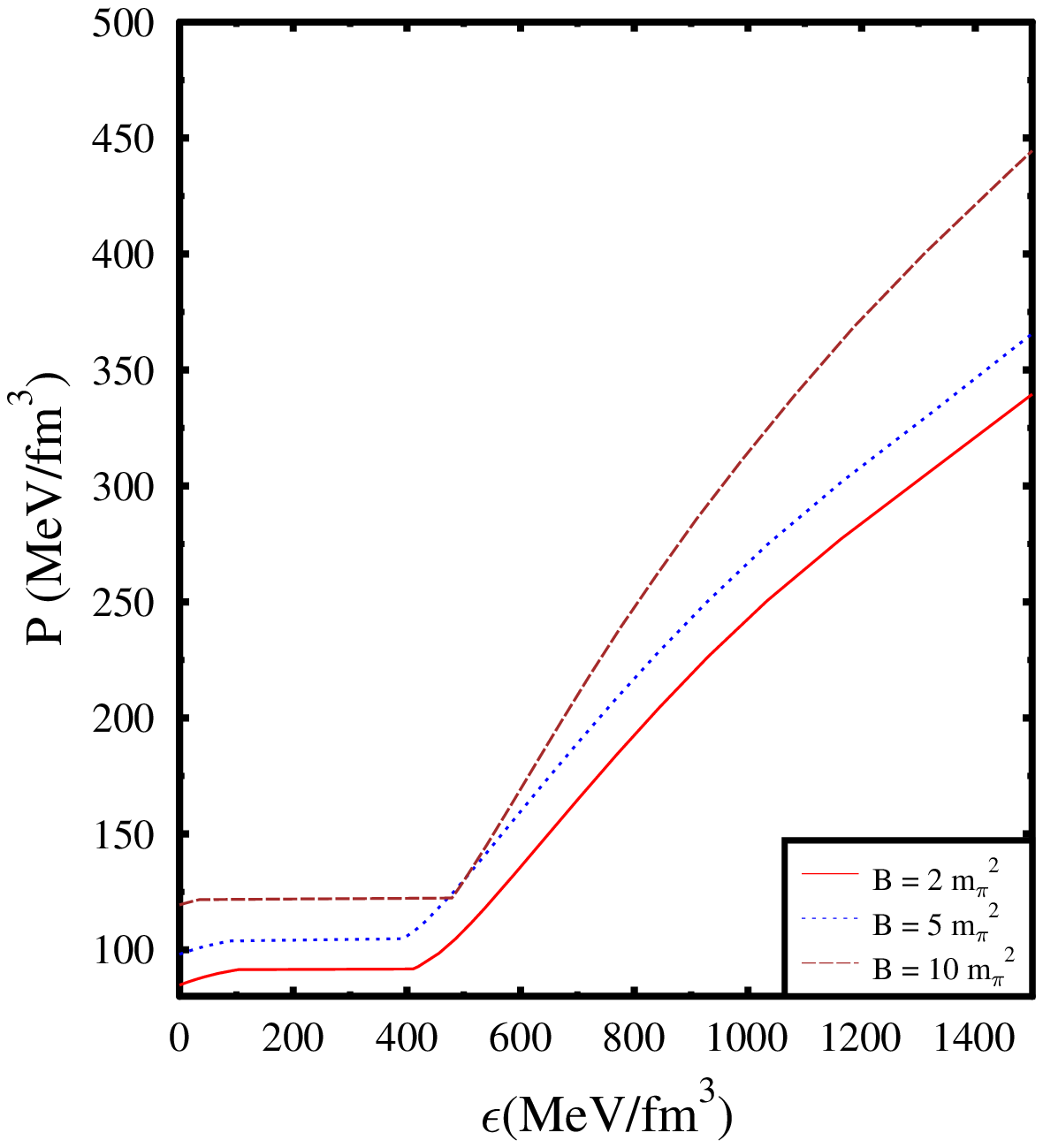}
\caption{Equation of state for charge neutral matter at T=40 MeV
for different magnetic field.}
\label{fig10}
\end{figure}

Let us note that in the equation of state that we have plotted in Fig.\ref{fig8} 
and Fig.\ref{fig10}, the pressure here corresponds to the thermodynamic pressure
i.e. negative of the thermodynamic potential given in Eq.(\ref{thpot1}). However,
in presence of magnetic field, the hydrodynamic pressure can be highly anisotropic
\cite{canuto,armendirk,ferrer3} when there is significant magnetization of the matter. 
The pressure in the direction of the field $P_\parallel$ is the thermodynamic
pressure $P=-\Omega$ as defined in Eq.(\ref{thpot1}). On the other hand, the pressure
$P_\perp$ in the transverse direction of the applied magnetic field is given by 
$P_\perp=P-MB$ \cite{armendirk}. Here, $M=-\partial\Omega/\partial B$ is the 
magnetization of the system. Using the thermodynamic potential expression given in Eq.(\ref{thpot1}),
this can be written as
\be
M=M_{med}+M_{field}+M_c
\label{mag}
\ee
where, $M_{med}$ , the contribution the magnetization from the medium which at
zero temperature is given by
\be
M_{med}=-\frac{\partial\Omega_{med}}{\partial B}=\frac{N_c}{4\pi^2}
\sum_{n,i}\alpha_n|q_i|\left[\mu_ip_{zmax}^i-(A_n^2+2n|q_i|B)
\log\frac{\mu_i+p_{zmax}^i}{A_n}\right]
\label{magmed}
\ee
where, we have abbreviated $A_n=\sqrt{M^2+2n|q_i|B}$. $M_{field}$ is the 
contribution from the
field part of the thermodynamic potential $\Omega_{field}$  given as
\be
M_{field}=-\frac{\partial\Omega_{field}}{\partial B}=\sum_i q_i^2B\left[\frac{\log x_i}{12}-\frac{1}{24}+x_i^3I_1(x_i)\right],
\label{magfield}
\ee
where,
\be
I_1(x)=\frac{1}{\pi}\int y
\frac{2\arctan (y)+y\log(1+y^2)}{(\exp(2\pi x y)-1)(1-\exp(-2\pi y)} dy
\ee
We might note here that $\Omega_{field}$ term originates from the effect of magnetic field
on the Dirac sea so that we can recognize that $M_{field}$ is due the magnetization
of the Dirac sea.

Finally, $M_c$ in Eq.(\ref{mag}) is the contribution to the magnetization
arising from the last two terms of the thermodynamic potential in Eq.(\ref{thpot1})
and is given as
\be
M_c=-4G\sum_iI_i\frac{\partial I_i}{\partial B}-2 K\sum_{i\neq j\neq k} I_iI_j
\frac{\partial I_k}{\partial B}.
\ee
Here, $\frac{\partial I_i}{\partial B}$ is the derivative of the 
quark condensate (-$\langle\bar\psi_i\psi\rangle$) with respect to the magnetic field
given as
$$\frac{\partial I_i}{\partial B}=\frac{\partial I_{med}^i}{\partial B}+
\frac{\partial I_{field}^i}{\partial B},$$
where, the contribution from the medium at zero temperature is
\be
\frac{\partial I_{med}^i}{\partial B}=\sum_n\frac{N_c\alpha_n}{2\pi^2}\left[\log\frac{
p_{zmax}^i+\mu_i}{A_n} -\frac{n|q_i|B}{A_n^2}\frac{\mu_i}{p_{zmax}^i}\right],
\ee
and, the field contribution from the condensate to magnetization
\be
\frac{\partial I_{field}^i}{\partial B}=\frac{N_c}{2\pi^2}\left[\log\Gamma(x_i)+
\frac{1}{2}\log\frac{x_i}{2\pi}+x_i-x_i\Psi^0(x_i)-\frac{1}{2}\right],
\ee
where, as defined earlier $x_i=(M_i^2/2|q_i|B)$ and
 $\Psi^0 (x)=\frac{\Gamma'(x)}{\Gamma(x)}$ is the logarithmic 
derivative of the Gamma function.

\begin{figure}
\includegraphics[width=8cm,height=8cm]{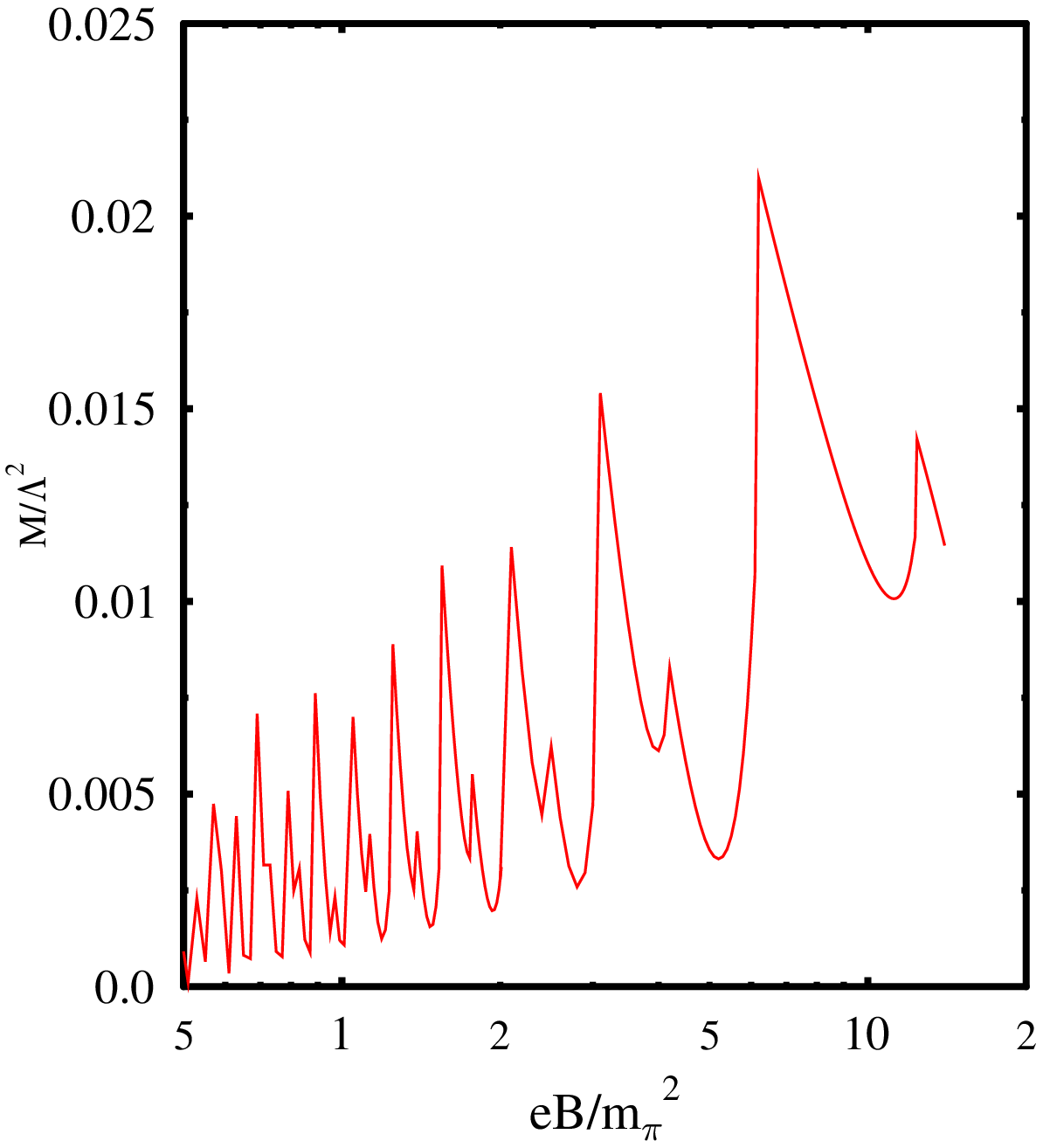}
\caption{Magnetization in units of $\Lambda^2$
as a function of magnetic field. The magnetic field in units of
$m_\pi^2$ is plotted in a logarithmic scale.
 We have taken here T=0 and $\mu_u=0.4GeV=\mu_d$.  }
\label{figmag}
\end{figure}

The resulting magnetization at T=0
 is plotted in Fig.\ref{figmag} for $\mu_u=0.4GeV=\mu_d$. 
The magnetization 
exhibits rapid de Hass-van Alphen oscillations. The irregularity in the oscillation
is due to the unequal masses of the three quarks which are calculated self consistently
using the gap equation. Unlike in Ref.\cite{armendirk}, the magnetization does not 
become constant even after the all the quarks are in the lowest Landau level. This
is due to the effect of contribution of magnetization from the Dirac sea which is
included here along with the Fermi sea contribution given by $M_{medium}$.

\begin{figure}
\includegraphics[width=8cm,height=8cm]{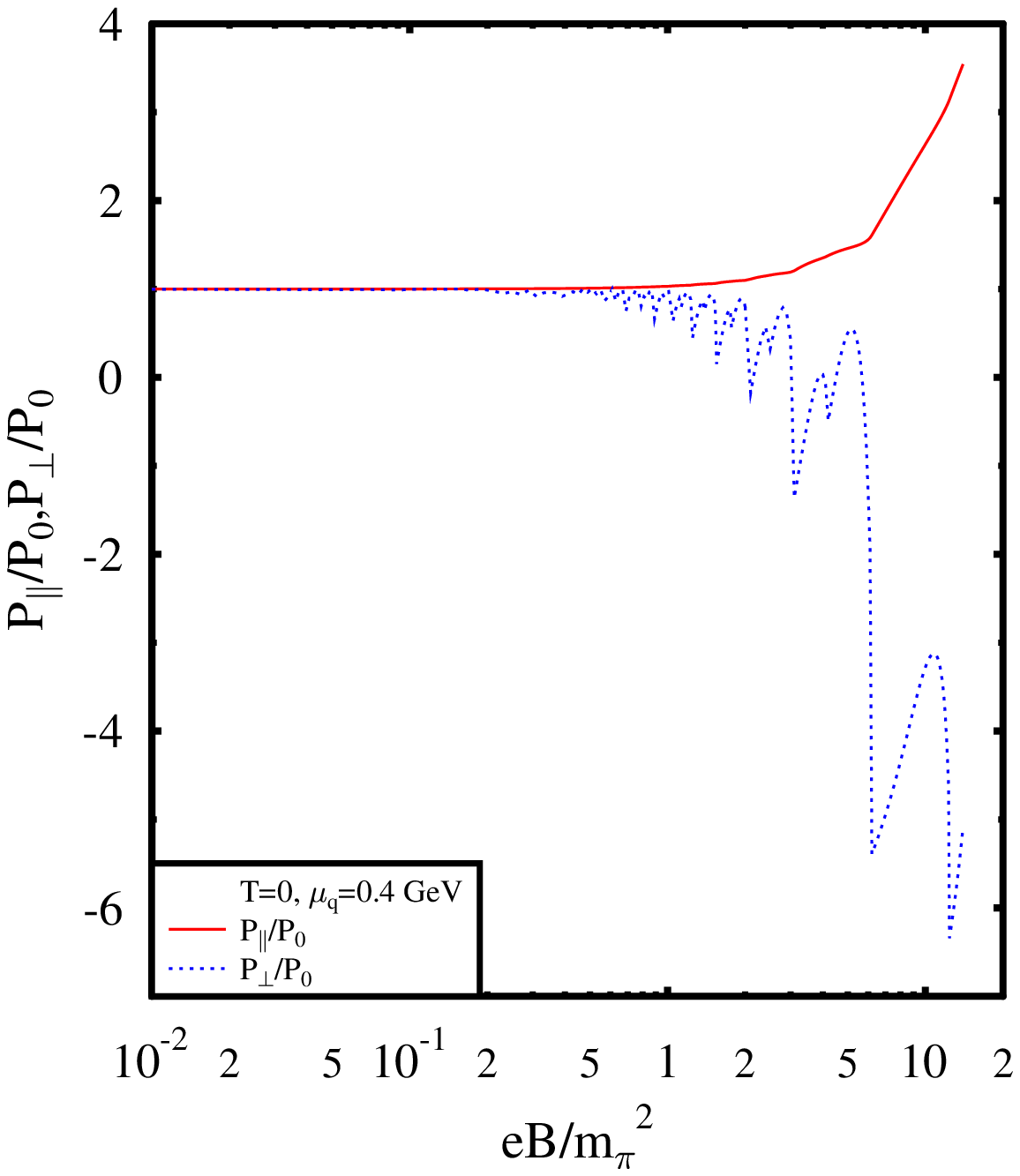}
\caption{The parallel $P_\parallel$ and transverse $P_\perp$ pressures of 
strange quark matter as functions of the magnetic field in units of pressure
$P_0$ for zero magnetic field. Magnetic field in units of $m_\pi^2$
is plotted in a logarithmic scale.
We have taken here T=0 and $\mu_u=0.4GeV=\mu_d$.  }
\label{figmag1}
\end{figure}

The transverse and the longitudinal pressure for the system is plotted in Fig.\ref{figmag1}.
Here, we have taken T=0 and $\mu=400$MeV. The oscillatory behavior of the magnetization
is reflected in the transverse pressure.
The two pressures start to differ significantly for magnetic field 
strengths of about $eB=m_\pi^2$
which corresponds to about $10^{18}$ Gauss. Such field induced 
anisotropy in pressure is qualitatively 
similar as in Ref. \cite{armendirk}, where, the anisotropic
properties of transport coefficients for strange quark matter were considered. 
While considering neutron star structure,
one has to also include the free field energy $\frac{1}{2}B^2$ to the total energy
and pressure. This term adds to the parallel and the transverse pressure 
with different signs \cite{ferrer3}. This can make the pressure in the transverse 
direction negative leading to mechanical instability \cite{ferrer3}. While studying
structural properties of compact astrophysical objects 
endowed with magnetic fields such splitting of the pressure in to the parallel 
and perpendicular direction need to be taken into account as this
can effect the structure and geometry of the star.

Finally we end this section with a comment regarding the axial fermion current density
induced at finite chemical potential. From Eq.s(\ref{j35p}) and Eq.(\ref{them}),
\be
\langle j_5^{i3}\rangle=
\frac{N_c|q_i|B}{(2\pi)^2}\int{dp_z\bigg[
\frac{1}{\exp(\sqrt{p_z^2+M_i^2}-\mu_i)}
-\frac{1}{\exp(\sqrt{p_z^2+M_i^2}+\mu_i)}}\bigg].
\label{j35pp}
\ee
Thus, although the lowest Landau level contributes to the above expectation value,
because of its dependence on the constituent quark mass parameter $M_i$, the effects
of all the higher Landau levels are implicitly there in Eq.(\ref{j35pp}) as 
the constituent masses here are calculated self consistently using 
Eq.(\ref{gapeq}) and Eq.(\ref{Iif}). Further, because of dependence on the
constituent quark mass the axial quark current density expectation value  also depends
upon the coupling in a nonperturbative manner \cite{igormag,hongmag}.

\section{summary}
We have analyzed here the ground state structure for chiral symmetry breaking 
in presence of strong magnetic field. The methodology
uses an explicit variational construct for the ground state in terms of
quark-antiquark pairing. A nice feature of the approach is that the 
four component quark field operator in presence of magnetic field could get expressed
in terms of the ansatz functions that occurs for the description of the ground state.
Apart from the methodology being new, we also have new results. Namely,
the present investigations have been done in a three flavor NJL model along 
with a flavor mixing six quark determinant interaction at finite temperature and density
and fields within the same framework. In that sense it generalizes the
two flavor NJL model considered in Ref.\cite{boomsma} for both finite temperature
and density.
The gap functions and 
the thermal distribution functions could be determined self consistently for 
given values of the temperature, the quark chemical potential and the 
strength of magnetic field.
At zero baryon density and high temperature, the qualitative feature
of chiral transition remains a cross over transition even
for magnetic field strength eB=10m$_\pi^2$. The magnetic catalysis of
chiral symmetry breaking is also observed.

At finite densities, the effects of Landau quantization is more dramatic. The order
parameter shows oscillation similar to the de Hass van Alphen effect for 
magnetization in metals. However, in the present case of dense quark matter, 
the mass of the quark itself is
dependant on the strength of magnetic fields which leads to a non periodic oscillation
of the order parameter. Although the critical chemical potential, $\mu_c$, for chiral
transition consistently decreases with
increase in the strength of the magnetic field, the corresponding density 
increases
with the magnetic field strength. Imposition of electrical charge neutrality
condition for the quark matter increases the value for $\mu_c$. Since 
the mass of the strange quark plays an important role in maintaining the
charge neutrality condition, this in turn affects the chiral restoration transition
in quark matter. The presence of nonzero magnetic field appears to make 
the equation of state stiffer.
Further, the pressure could be anisotropic if the magnetization of
the matter is significant. Within the model, this anisotropy starts to become 
relevant for field strengths around $10^{18}$ Gauss. While considering the structural
properties of astro physical compact objects having magnetic fields this
anisotropy in the equation of state should be taken into account as it 
can affect the geometry and structure of the star.

We have considered here quark-antiquark pairing in our ansatz 
for the ground state which is homogeneous with zero total momentum 
as in Eq.(\ref{ansatz}). However, it is possible that 
the condensate could be spatially non-homogeneous with a net total 
momentum \cite{dune,frolov,nickel}. Further, one could  include the effect of
deconfinement transition by generalizing the present model to Polyakov
loop NJL models for three flavors to investigate the inter relationship
of deconfinement and the chiral transition in presence of strong fields
for  the three flavor case
considered here\cite{gato}. This will be particularly important
for finite temperature calculations. At finite density and small
temperatures, the ansatz can be generalized to include the
diquark condensates in presence of magnetic field \cite{digal,iran,ferrerscmag}.
Some of these calculations are
in progress and will be reported elsewhere.

\acknowledgments
We would like to thank the organizers of WHEEP-10 where part of the present work
was completed.
We would like to thank V. Tiwari, U.S. Gupta, R. Roy and Sayantan Sharma
for discussions. One of the authors (AM) would like to acknowledge support from
Department of Science and Technology, Governemnt of India 
(Project No. SR/S2/HEP-21/2006). She would also
like to thank Frankfurt Institute for Advanced Studies (FIAS) for warm hospitality, 
where the present work was completed and Alexander von Humboldt Foundation,
 Germany for financial support.

\appendix
\section{Spinors in a constant magnetic field}
Here we derive the solutions for the spinors for a relativistic charged 
particle
in presence of an external constant magnetic field and the field operator expansion
for the corresponding fermion field. We shall take the direction of the
magnetic field to be in the z-direction. We choose the corresponding gauge
field $A_{\mu}=(0,0,Bx,0)$. The Dirac equation in presence of the
uniform magnetic field is then written as

\begin{equation}
 i\frac{\partial\psi}{\partial t} = (\vec\alpha\cdot\vec\Pi + \beta
m)\psi,
\label{DirEq}
\end{equation}
where, $\vec\Pi = \zbf p - q\zbf A$ is the kinetic momentum of the
particle with electric charge $q$ in the presence of the magnetic field.

Let us first derive the positive energy solutions $U(x)$ of Eq.(\ref{DirEq}).
For stationary solution of energy $E$ we choose $U (x)$ as
\be
 U(\zbf x,t) =
\left[\begin{array}{c}
\phi(\zbf x)\\ 
\chi(\zbf x)\\
\end{array}\right] e^{-iEt},
\label{uspin}
\ee
where $\phi(\zbf x)$ and $\chi(\zbf x)$ are the two component
spinors. Substitution this ansatz in Eq.(\ref{DirEq})  leads to

\be
 \chi(\vec x) = \frac{\vec\sigma\cdot\vec\Pi}{E+m}\phi(\vec x),
\label{chi}
\ee

so that eliminating $\chi$ in favor of $\phi$, leads to an equation 
for the latter as
\be
 (E^2 - m^2)\phi(\vec x) = (\vec\sigma\cdot\vec\Pi)^2\phi(\vec x).
\ee
Noting that $(\zbf \sigma\cdot\zbf\Pi)^2=\zbf\Pi^2-q\zbf\sigma\cdot\zbf B$. With
$\zbf B=(0,0,B)$ and with our choice of the gauge $A_\mu=(0,0,Bx,0)$, the above 
equation reduces to
\begin{equation}
 (E^2 - m^2)\phi(\zbf x) = [-\nabla^2+(qBx)^2-qB(\sigma_3+2xp_y)]\phi(\zbf x)
\label{eqphi}
\end{equation}
Next, recognizing the fact that in the RHS, the coordinates $y$ and $z$ do not occur
explicitly except for in the derivatives, we might assume the solution to be of the form
\begin{equation}
 \phi(\zbf x) = e^{i(p_y y+p_z z)}f(x)
\label{phi}
\end{equation}
where, $f(x)=f_\alpha u_\alpha$, $\alpha=\pm 1$ for spin up and spin down 
respectively with 
$u_{1}=\left(\begin{array}{c}
1\\
0\\
\end{array}\right)$; 
$u_{-1}=\left(\begin{array}{c}
0\\
1\\
\end{array}\right)$; so that $\sigma_3 f=\alpha f$. Using Eq.(\ref{phi}) in 
Eq.(\ref{eqphi}) we have,
\be
 \left[\frac{\partial^2}{\partial\xi^2}-\xi^2+a_\alpha\right]f_\alpha(\xi) = 0,
\label{eqfxi}
\end{equation}
where, we have introduced the dimensionless variables 
$\xi = \sqrt{|q|B}\left(x-\frac{p_y}{qB}\right)$ and  $a_\alpha = 
\frac{E^2-m^2-p_z^2+qB\alpha}{|q| B}$. Eq.(\ref{eqfxi}) is a special form of
Hermite differential equation, whose solutions exist for
 $a_\alpha = 2n+1$, $n=0,1,2,...$. This 
 gives the energy levels as
\begin{equation}
 E_{n\alpha}^2 = m^2+p_z^2+(2n+1)|q|B-qB\alpha.
\label{+veenergy}
\end{equation}
The solution of Eq.(\ref{eqfxi}) is 
\be
 f_\alpha(\xi) = c_ne^{-\frac{\xi^2}{2}}H_n(\xi) = I_n(\xi),
\label{inxi1}
\ee
Where $H_n(\xi)$ is the Hermite polynomial of the nth order, 
with the normalization
constant $c_n$ given by
\begin{equation*}
 c_n = \sqrt{\frac{\sqrt{|q|B}}{n!2^n\sqrt{\pi}}}.
\end{equation*}
The functions $I_n(\xi)$'s satisfy the completeness relation
\be
\sum_n I_n(\xi)I_n(\xi`)=|q|B\delta(\xi-\xi').
\label{complete}
\ee
Further using orthonormality condition for the Hermite polynomials, $I_n$'s are
normalized as
\be
\int d\xi I_n(\xi)I_m(\xi)=\sqrt{|q|B}\delta_{n,m},
\label{ortho}
\ee
$I_n(\xi)$'s are seen to satisfy the following relations
\begin{mathletters}
\be
\frac{\partial}{\partial x}I_n(\xi) = \sqrt{|q|B}[-\xi
I_n(\xi)+\sqrt{2n}I_{n-1}(\xi)] \ee
\be
2\xi I_n(\xi) = \sqrt{2n}I_{n-1}(\xi) + \sqrt{2(n+1)}I_{n+1}(\xi)
\label{recuri}
\ee
\end{mathletters}

Thus with the upper component $\phi(\zbf x)$ known from Eq.(\ref{phi}) and 
Eq.(\ref{inxi1}), the lower component $\chi(\zbf x)$ can be evaluated from
Eq.(\ref{chi}) using the relations Eq.(\ref{recuri}). This leads to the explicit
solutions for the positive energy spinors as $U(\zbf x,t)=U(n,\zbf p_{\omit x},x)
\exp(i\zbf p_{\omit x}\cdot \zbf x_{\omit x}-i\epsilon_n t)$ with
\begin{subequations}
\begin{eqnarray}
U_{\uparrow }(x,\vec p_{\omit x},n) &=& \frac{1}{\sqrt{2\epsilon_n(\epsilon_n+m)}}
\left(\begin{array}{c}
(\epsilon_n+m)\left(\Theta(q)I_n + \Theta(-q)I_{n-1}\right)\\
0\\
p_z\left(\Theta(q)I_n+\Theta(-q)I_{n-1}\right)\\
-i\sqrt{2n|q|B}\left(\Theta(q)I_{n-1}+\Theta(-q)I_{n}\right)\\
\end{array}\right) \\
 U_{\downarrow}(x,\vec p_{\omit x},n) &=& \frac{1}{\sqrt{2\epsilon_n(\epsilon_n+m)}}
\left(\begin{array}{c}
0 \\ 
(\epsilon_n+m)\left(\Theta(q)I_{n-1}+\Theta(-q)I_n\right) \\
i\sqrt{2n|q|B}\left(\Theta(q)I_n-\Theta(-q)I_{n-1}\right)
\\
-p_z\left(\Theta(q)I_n-\Theta(-q)I_{n-1}\right)\\
\end{array}\right). 
\label{Us}
\end{eqnarray}
\end{subequations}

In the above, we have defined $\epsilon_n=\sqrt{p_z^2+m^2+2 n|q|B}$ and further,
we have defined $I_{-1}=0$ while for nonnegative values of $n$, 
$I_n(\xi)$'s are given by Eq.(\ref{inxi1}).

In an identical manner, one can obtain the solutions for the antiparticles and 
the solution can be written as $V(\zbf x,t)=V(x,\zbf p_{\omit x},n)exp(-i\zbf p_{\omit x}
\cdot \zbf x_{\omit x }+i\epsilon_n t)$ with
\begin{subequations}
\begin{eqnarray}
V_{\uparrow}(x,-\zbf p_{\omit x},n)& =& \frac{1}{\sqrt{2\epsilon_n(\epsilon_n+m)}}
\left(\begin{array}{c}
\sqrt{2n|q|B}
\left(\Theta(q)I_n-\Theta(-q)I_{n-1}\right)
\\ 
ip_z
\left(\Theta(q)I_{n-1}+\Theta(-q)I_{n}\right)
\\
0\\
i(\epsilon_n+m)
\left(\Theta(q)I_{n-1}+\Theta(-q)I_{n}\right)
\\
\end{array}\right),\\
V_{\downarrow}(x,-\vec p_{\omit x},n) &=& \frac{1}{\sqrt{2\epsilon_n(\epsilon_n+m)}}
\left[\begin{array}{c}
ip_z
\left(\Theta(q)I_{n}+\Theta(-q)I_{n-1}\right)
\\ 
\sqrt{2n|q|B}
\left(\Theta(q)I_{n-1}-\Theta(-q)I_{n-1}\right)
 \\
-i(\epsilon_n+m)
\left(\Theta(q)I_{n}+\Theta(-q)I_{n-1}\right)
\\
0\\
\end{array}\right).
\label{Vs}
\end{eqnarray}
\end{subequations}

The spinors are normalized as
\be
\int dx U_r(x,\zbf p_{\omit x},n)^\dagger U_s(x,\zbf p_{\omit x},m)
=\delta_{n,m}\delta_{r,s}
=\int dx V_r(x,\zbf p_{\omit x},n)^\dagger V_s(x,\zbf p_{\omit x},m)
\label{normspinor}
\ee
These spinors are used in Eq.(\ref{psiex}) for expansion of the field operators
in the momentum space. 

\def \ltg{R.P. Feynman, Nucl. Phys. B 188, 479 (1981); 
K.G. Wilson, Phys. Rev. \zbf  D10, 2445 (1974); J.B. Kogut,
Rev. Mod. Phys. \zbf  51, 659 (1979); ibid  \zbf 55, 775 (1983);
M. Creutz, Phys. Rev. Lett. 45, 313 (1980); ibid Phys. Rev. D21, 2308
(1980); T. Celik, J. Engels and H. Satz, Phys. Lett. B129, 323 (1983)}

\def\berges {J. Berges, K. Rajagopal, {\NPB{538}{215}{1999}}.}% Nucl. Phys. B538, 215, (1999).}
\def \svz {M.A. Shifman, A.I. Vainshtein and V.I. Zakharov,
Nucl. Phys. B147, 385, 448 and 519 (1979);
R.A. Bertlmann, Acta Physica Austriaca 53, 305 (1981)}

\def \spmbst {S.P. Misra, Phys. Rev. D35, 2607 (1987)}

\def \hmgrnv { H. Mishra, S.P. Misra and A. Mishra,
Int. J. Mod. Phys. A3, 2331 (1988)}

\def \snss {A. Mishra, H. Mishra, S.P. Misra
and S.N. Nayak, Phys. Lett 251B, 541 (1990)}

\def \amqcd { A. Mishra, H. Mishra, S.P. Misra and S.N. Nayak,
Pramana (J. of Phys.) 37, 59 (1991). }
\def\qcdtb{A. Mishra, H. Mishra, S.P. Misra 
and S.N. Nayak, Z.  Phys. C 57, 233 (1993); A. Mishra, H. Mishra
and S.P. Misra, Z. Phys. C 58, 405 (1993)}
% pure qcd at zero and finite temperature and finite baryon densities

\def \spmtlk {S.P. Misra, Talk on {\it `Phase transitions in quantum field
theory'} in the Symposium on Statistical Mechanics and Quantum field theory, 
Calcutta, January, 1992, hep-ph/9212287}

\def \hmspmnjl {H. Mishra and S.P. Misra, 
{\PRD{48}{5376}{1993}.}}
%Phys. Rev. D {\bf 48},5376 (1993)}

\def \hmqcd {A. Mishra, H. Mishra, V. Sheel, S.P. Misra and P.K. Panda,
hep-ph/9404255 (1994)}

\def \amcrl {A. Mishra, H. Mishra and S.P. Misra, Z. Phys. C 57, 241 (1993)}

\def \higgs { S.P. Misra, in {\it Phenomenology in Standard Model and Beyond}, 
Proceedings of the Workshop on High Energy Physics Phenomenology, Bombay,
edited by D.P. Roy and P. Roy (World Scientific, Singapore, 1989), p.346;
A. Mishra, H. Mishra, S.P. Misra and S.N. Nayak, Phys. Rev. D44, 110 (1991)}

\def \nmtr {A. Mishra, 
H. Mishra and S.P. Misra, Int. J. Mod. Phys. A5, 3391 (1990); H. Mishra,
 S.P. Misra, P.K. Panda and B.K. Parida, Int. J. Mod. Phys. E 1, 405, (1992);
 {\it ibid}, E 2, 547 (1993); A. Mishra, P.K. Panda, S. Schrum, J. Reinhardt
and W. Greiner, to appear in Phys. Rev. C}
 % nuclear matter

\def \dtrn {P.K. Panda, R. Sahu and S.P. Misra, 
Phys. Rev C45, 2079 (1992)}
\def\hurwitz{E. Elizalde, J. Phys. {\bf A}:Math. Gen. 18,1637 (1985).}

\def \qcd {G. K. Savvidy, Phys. Lett. 71B, 133 (1977);
S. G. Matinyan and G. K. Savvidy, Nucl. Phys. B134, 539 (1978); N. K. Nielsen
and P. Olesen, Nucl.  Phys. B144, 376 (1978); T. H. Hansson, K. Johnson,
C. Peterson Phys. Rev. D26, 2069 (1982)}
% qcd vacuum considered earlier

\def \cornwal {J.M. Cornwall, Phys. Rev. D26, 1453 (1982)}
\def\aichlin {F. Gastineau, R. Nebauer and J. Aichelin,
{\PRC{65}{045204}{2002}}.}
% Phys. Rev. C65, 045204 (2002).}

\def \mndglv {J. E. Mandula and M. Ogilvie, Phys. Lett. 185B, 127 (1987)}

\def \schwinger {J. Schwinger, Phys. Rev. 125, 1043 (1962); ibid,
127, 324 (1962)}

\def \schutte {D. Schutte, Phys. Rev. D31, 810 (1985)}

\def \amspm {A. Mishra and S.P. Misra, 
{\ZPC{58}{325}{1993}}.}
%Z. Phys. C 58, 325 (1993)}

\def \gft{ For gauge fields in general, see e.g. E.S. Abers and 
B.W. Lee, Phys. Rep. 9C, 1 (1973)}

\def \gribov {V.N. Gribov, Nucl. Phys. B139, 1 (1978)}

\def \spm78 {S.P. Misra, Phys. Rev. D18, 1661 (1978); {\it ibid}
D18, 1673 (1978)} 

\def \lopr {A. Le Youanc, L.  Oliver, S. Ono, O. Pene and J.C. Raynal, 
Phys. Rev. Lett. 54, 506 (1985)}

\def \spphi {S.P. Misra and S. Panda, Pramana (J. Phys.) 27, 523 (1986);
S.P. Misra, {\it Proceedings of the Second Asia-Pacific Physics Conference},
edited by S. Chandrasekhar (World Scientific, 1987) p. 369}

\def\spmdif {S.P. Misra and L. Maharana, Phys. Rev. D18, 4103 (1978); 
    S.P. Misra, A.R. Panda and B.K. Parida, Phys. Rev. Lett. 45, 322 (1980);
    S.P. Misra, A.R. Panda and B.K. Parida, Phys. Rev. D22, 1574 (1980)}

\def \spmvdm {S.P. Misra and L. Maharana, Phys. Rev. D18, 4018 (1978);
     S.P. Misra, L. Maharana and A.R. Panda, Phys. Rev. D22, 2744 (1980);
     L. Maharana,  S.P. Misra and A.R. Panda, Phys. Rev. D26, 1175 (1982)}

\def\spmthr {K. Biswal and S.P. Misra, Phys. Rev. D26, 3020 (1982);
               S.P. Misra, Phys. Rev. D28, 1169 (1983)}

\def \spmstr { S.P. Misra, Phys. Rev. D21, 1231 (1980)} 

\def \spmjet {S.P. Misra, A.R. Panda and B.K. Parida, Phys. Rev Lett. 
45, 322 (1980); S.P. Misra and A.R. Panda, Phys. Rev. D21, 3094 (1980);
  S.P. Misra, A.R. Panda and B.K. Parida, Phys. Rev. D23, 742 (1981);
  {\it ibid} D25, 2925 (1982)}

\def \arpftm {L. Maharana, A. Nath and A.R. Panda, Mod. Phys. Lett. 7, 
2275 (1992)}

\def \van {R. Van Royen and V.F. Weisskopf, Nuov. Cim. 51A, 617 (1965)}

\def \rchpi {S.R. Amendolia {\it et al}, Nucl. Phys. B277, 168 (1986)}
% pion charge radius (experimental) .66+-.01 fm; rch2=11.22 Gev-2.

\def \chrl{ Y. Nambu, {\PRL{4}{380}{1960}};
%Phys. Rev. Lett. \zbf 4, 380 (1960);
A. Amer, A. Le Yaouanc, L. Oliver, O. Pene and
J.C. Raynal,{\PRL{50}{87}{1983a}};{\em ibid}
{\PRD{28}{1530}{1983}};
% Phys. Rev. Lett.\zbf  50, 87 (1983);
%ibid, Phys. Rev.\zbf  D28, 1530 (1983); 
M.G. Mitchard, A.C. Davis and A.J.
MAacfarlane, {\NPB{325}{470}{1989}};
%Nucl. Phys. \zbf B325, 470 (1989);
B. Haeri and M.B. Haeri,{\PRD{43}{3732}{1991}};
% Phys. Rev.\zbf  D43, 3732 (1991); 
V. Bernard,{\PRD{34}{1604}{1986}};
% Phys. Rev.\zbf  D34, 1601 (1986);
 S. Schramm and
W. Greiner, Int. J. Mod. Phys. \zbf E1, 73 (1992), 
J.R. Finger and J.E. Mandula, Nucl. Phys. \zbf B199, 168 (1982),
S.L. Adler and A.C. Davis, Nucl. Phys.\zbf  B244, 469 (1984),
S.P. Klevensky, Rev. Mod. Phys.\zbf  64, 649 (1992).}
\def\klevansky{S.P. Klevansky, Rev. Mod. Phys.\zbf  64, 649 (1992).}

\def \spmijp { S.P. Misra, Ind. J. Phys. 61B, 287 (1987)}

\def \feynman {R.P. Feynman and A.R. Hibbs, {\it Quantum mechanics and
path integrals}, McGraw Hill, New York (1965)}

\def \glstn{ J. Goldstone, Nuov. Cim. \zbf 19, 154 (1961);
J. Goldstone, A. Salam and S. Weinberg, Phys. Rev. \zbf  127,
965 (1962)}

\def \anderson {P.W. Anderson, Phys. Rev. \zbf {110}, 827 (1958)}

\def \nambu{ Y. Nambu, Phys. Rev. Lett. \zbf 4, 380 (1960)}

\def\donogh {J.F. Donoghue, E. Golowich and B.R. Holstein, {\it Dynamics
of the Standard Model}, Cambridge University Press (1992)}

\def\satz {T. Matsui and H. Satz, Phys. Lett. B178, 416 (1986)}

\def\cps {C. P. Singh, Phys. Rep. 236, 149 (1993)}

\def\prliop {A. Mishra, H. Mishra, S.P. Misra, P.K. Panda and Varun
Sheel, Int. J. of Mod. Phys. E 5, 93 (1996)}

\def\hmcor {V. Sheel, H. Mishra and J.C. Parikh, Phys. Lett. B382, 173
(1996); {\it biid}, to appear in Int. J. of Mod. Phys. E}
\def\cort { V. Sheel, H. Mishra and J.C. Parikh, Phys. ReV D59,034501 (1999);
{\it ibid}Prog. Theor. Phys. Suppl.,129,137, (1997).}
% J. Phys. G23,143, (1997).}

\def\surcor {E.V. Shuryak, Rev. Mod. Phys. 65, 1 (1993)} 

\def\stevenson {A.C. Mattingly and P.M. Stevenson, Phys. Rev. Lett. 69,
1320 (1992); Phys. Rev. D 49, 437 (1994)}

\def\mac {M. G. Mitchard, A. C. Davis and A. J. Macfarlane,
 Nucl. Phys. B 325, 470 (1989)} 
\def\tfd
 {H.~Umezawa, H.~Matsumoto and M.~Tachiki {\it Thermofield dynamics
and condensed states} (North Holland, Amsterdam, 1982) ;
P.A.~Henning, Phys.~Rep.253, 235 (1995).}
\def\amph4{Amruta Mishra and Hiranmaya Mishra,
{\JPG{23}{143}{1997}}.}

\def \neglecor{M.-C. Chu, J. M. Grandy, S. Huang and 
J. W. Negele, Phys. Rev. D48, 3340 (1993);
ibid, Phys. Rev. D49, 6039 (1994)}

\def\revdata {Particle Data Group, Phys. Rev. D 50, 1173 (1994)}

\def\sinp {S.P. Misra, Indian J. Phys., {\bf 70A}, 355 (1996)}
\def\hmparikh{H. Mishra and J.C. Parikh, {\NPA{679}{597}{2001}.}}
% Nucl. Physics A679, 597 (2001).}
\def\krisch {M. Alford and K. Rajagopal, JHEP 0206,031,(2002)}
\def\reddy {A.W. Steiner, S. Reddy and M. Prakash,
{\PRD{66}{094007}{2002}.}}
\def\hmam {Amruta Mishra and Hiranmaya Mishra,
{\PRD{69}{014014}{2004}.}}
% Phys. Rev. D66, 094007, 2002}
\def\hmampp {Amruta Mishra and Hiranmaya Mishra,
in preparation.}
\def\bryman {D.A. Bryman, P. Deppomier and C. Le Roy, Phys. Rep. 88,
151 (1982)}
\def\thooft {G. 't Hooft, Phys. Rev. D 14, 3432 (1976); D 18, 2199 (1978);
S. Klimt, M. Lutz, U. Vogl and W. Weise, Nucl. Phys. A 516, 429 (1990)}
\def\alkz { R. Alkofer, P. A. Amundsen and K. Langfeld, Z. Phys. C 42,
199(1989), A.C. Davis and A.M. Matheson, Nucl. Phys. B246, 203 (1984).}
\def\sarah {T.M. Schwartz, S.P. Klevansky, G. Papp,
{\PRC{60}{055205}{1999}}.}
% Phys. Rev. C60, 055205 (1999)}
\def\wil{M. Alford, K.Rajagopal, F. Wilczek, {\PLB{422}{247}{1998}};
%Phys. Lett. B422,247(1998), 
{\it{ibid}}{\NPB{537}{443}{1999}}.}
%Nucl. Phys. B537,443 (1999).}
\def\sursc{R.Rapp, T.Schaefer, E. Shuryak and M. Velkovsky,
{\PRL{81}{53}{1998}};{\it ibid}{\AP{280}{35}{2000}}.}
% Phys. Rev. Lett.  81, 53(1998),{\it {ibid}},Ann. Phys. 280, 35, 2000.}
\def\pisarski{
D. Bailin and A. Love, {\PR{107}{325}{1984}},
%Phys. Rep. 107 (1984) 325,
D. Son, {\PRD{59}{094019}{1999}}; 
%Phys. Rev. D59 (1999) 094019,
T. Schaefer and F. Wilczek, {\PRD{60}{114033}{1999}};
%Phys. Rev. D60 (1999) 114033,
D. Rischke and R. Pisarski, {\PRD{61}{051501}{2000}}, 
%Phys. Rev. D61 (2000) 051501,
D. K. Hong, V. A. Miransky, 
I. A. Shovkovy, L.C. Wiejewardhana, {\PRD{61}{056001}{2000}}.}
% Phys. Rev. D61 (2000) 056001.}
\def\leblac {M. Le Bellac, {\it Thermal Field Theory}(Cambridge, Cambridge University
Press, 1996).}
\def\bcs{A.L. Fetter and J.D. Walecka, {\it Quantum Theory of Many
particle Systems} (McGraw-Hill, New York, 1971).}
\def\alexander{Aleksander Kocic, Phys. Rev. D33, 1785,(1986).}
\def\bubmix{F. Neumann, M. Buballa and M. Oertel,
{\NPA{714}{481}{2003}.}}
% Nucl. Phys. A714:481-501,2003.}
\def\kunihiro{M. Kitazawa, T. Koide, T. Kunihiro, Y. Nemeto,
{\PTP{108}{929}{2002}.}}
% Prog.  theo. Phys. 108, 929, 2002.}
\def\igor{Igor Shovkovy, Mei Huang, {\PLB{564}{205}{2003}}.}
\def\prasanth{P. Jaikumar and M. Prakash,{\PLB{516}{345}{2001}}.}
\def\igorr{Mei Huang, Igor Shovkovy, {\NPA{729}{835}{2003}}.}
%Phys. Rev. D67, 103004, 2003.}
\def\abrikosov{A.A. Abrikosov, L.P. Gorkov, Zh. Eskp. Teor.39, 1781,
1960}
\def\krischprl{M.G. Alford, J. Berges and K. Rajagopal,
 {\PRL{84}{598}{2000}.}}
%Phys. Rev. Lett. 84, 598, 2000.}
\def\hatmampp{A. Mishra and H.Mishra, in preparation}
\def\blaschke{D. Blaschke, M.K. Volkov and V.L. Yudichev,
{\EPJA{17}{103}{2003}}.}
% Eur. Phys. J. A17,103, 2003}
\def\mei{M. Huang, P. Zhuang, W. Chao,
{\PRD{65}{076012}{2002}}}
% Phys. Rev D65, 076012, 2002}
\def\bubnp{M. Buballa, M. Oertel,
{\NPA{703}{770}{2002}}.}
\def\sarma{G. Sarma, J. Phys. Chem. Solids 24,1029 (1963).}
% Nucl. Phys. A703, 770 (2002);
\def\ebert {D. Ebert, H. Reinhardt and M.K. Volkov,
Prog. Part. Nucl. Phys.{\bf 33},1, 1994.}
\def\ebertmag {D. Ebert, K.G. Klimenko, M.A. Vdovichenko and A.S. Vshivtsev,
{\PRD{61}{025005}{1999}.}}
\def\ferrerscmag{E.J. Ferrer, V. de la Incera and C. Manuel, {\PRL{95}{152002}{2005}};
E.J. Ferrer and V. de la Incera, {\PRL{97}{122301}{2006}}; E.J. Ferrer and
V. de la Incera, {\PRD{76}{114012}{2007}}.}
\def\manfer{E.J. Ferrer, V. de la Incera and C. Manuel, Nucl. Phys.
 B{\bf 747},88 (2006).}
\def\klimenkoplb{K.G. Klimenko and V. Ch. Zhukovsky,{\PLB{665}{352}{2008}}.}
\def\andreasrebhan{F. Preis, A. Rebhan and A. Schmitt, JHEP 1103(2011),033.}
\def\canuto{ V. Canuto and J. Ventura, Fundam. Cosmic Phys.{\bf 2}, 203 (1977).}
\def\ferrer3{E. J. Ferrer, V. Incera and J. P. Keith, I. Portillo and
P. Springsteen, {\PRC{82}{065802}{2010}}.}
\def\rehberg{ P. Rehberg, S.P. Klevansky and J. Huefner,
{\PRC{53}{410}{1996}.}}
\def\lutz{M. Lutz, S. Klimt, W. Weise,{\NPA{542}{521}{1992}.}}
% Phys. Rev. C53,410 (1996).}
%\begin{references}
\def\rapid{B. Deb, A.Mishra, H. Mishra and P. Panigrahi,
Phys. Rev. A {\bf 70},011604(R), 2004.}
\def\kriscfl{M. Alford, C. Kouvaris, K. Rajagopal, Phys. Rev. Lett.
{\bf 92} 222001 (2004), arXiv:hep-ph/0406137.}
\def\shovris{S.B. Ruester, I.A. Shovkovy and D.H. Rischke,
arXiv:hep-ph/0405170.}
\def\spmindianj{S. P. Misra, Indian J. Phys. {\bf 70A}, 355 (1996).}
\def\kausik{K. Bhattacharya,arXiv:0705.4275[hep-th]; M. deJ. Aguiano-Galicia,
A. Bashir and A. Raya,{\PRD{76}{127702}{2007}}.}
\def\krisaug{K. Fukushima, C. Kouvaris and K. Rajagopal, arxiv:hep-ph/0408322}.
\def\wilczek{W.V. Liu and F. Wilczek,{\PRL{90}{047002}{2003}},E. Gubankova,
W.V. Liu and F. Wilczek, {\PRL{91}{032001}{2003}.}}
\def\review{For reviews see K. Rajagopal and F. Wilczek,
arXiv:hep-ph/0011333; D.K. Hong, Acta Phys. Polon. B32,1253 (2001);
M.G. Alford, Ann. Rev. Nucl. Part. Sci 51, 131 (2001); G. Nardulli,
Riv. Nuovo Cim. 25N3, 1 (2002); S. Reddy, Acta Phys Polon.B33, 4101(2002);
T. Schaefer arXiv:hep-ph/0304281; D.H. Rischke, Prog. Part. Nucl. Phys. 52,
197 (2004); H.C. Ren, arXiv:hep-ph/0404074; M. Huang, arXiv: hep-ph/0409167;
I. Shovkovy, arXiv:nucl-th/0410191.}
\def\kunihiroo{ M. Kitazawa, T. Koide, T. Kunihiro and Y. Nemoto,
{\PRD{65}{091504}{2002}}, D.N. Voskresensky, arXiv:nucl-th/0306077.}
\def\rupak{S.Reddy and G. Rupak, arXiv:nucl-th/0405054}
\def\ida{K. Iida and G. Baym,{\PRD{63}{074018}{2001}},
Erratum-ibid{\PRD{66}{059903}{2002}}; K. Iida, T. Matsuura, M. Tachhibana 
and T. Hatsuda, {\PRL{93}{132001}{2004}}; ibid,{arXiv:hep-ph/0411356}}
\def\chromo{Mei Huang and Igor Shovkovy,{\PRD{70}{051501}{2004}};
 {\em ibid}, {\PRD{70}{094030}{2004}}}
\def\steiner{A.W. Steiner, {\PRD{72}{054024}{2005}.}}
\def\andreaskris{K. Rjagopal and A. Schimitt{\PRD{73}{045003}{2006}.}}
\def\amhm5{A. Mishra and H. Mishra, {\PRD{71}{074023}{2005}.}}
\def\leupold{K. Schertler, S. Leupold and J. Schaffner-Bielich,
{\PRC{60}{025801}(1999).}}
\def\bubrep{Michael Buballa, Phys. Rep.{\bf 407},205, 2005.}
\def\hatkun{T. Hatsuda and T. Kunihiro, Phys. Rep.{\bf 247},221, 1994.}
\def\lkw{ M. Lutz, S. Klimt and W. Weise, Nucl Phys. {\bf A542}, 521, 1992.}
\def\ruester{S.B. Ruester, V.Werth, M. Buballa, I. Shovkovy, D.H. Rischke,
arXiv:nucl-th/0602018; S.B. Ruester, I. Shovkovy, D.H. Rischke,
{\NPA{743}{127}{2004}.}}
\def\larrywarringa{D.Kharzeev, L. McLerran and H. Warringa, {\NPA{803}{227}{2008}};
K.Fukushima, D. Kharzeev and H. Warringa,{\PRD{78}{074033}{2008}}.}
\def\skokov{V. Skokov, A. Illarionov and V. Toneev, Int. j. Mod. Phys. A {\bf 24}, 5925,
(2009).}
\def\dima{D. Kharzeev, Ann. of Physics, K. Fukushima, M. Ruggieri and R. Gatto,
{\PRD{81}{114031}{2010}}.}
\def\fraga{A.J. Mizher, M.N. Chenodub and E. Fraga,arXiv:1004.2712[hep-ph].}
\def\maglat{M.D'Elia, S. Mukherjee and F. Sanflippo,{\PRD{82}{051501}{2010}.}}
\def\fukushimaplb{K. Fukushima, M. Ruggieri and R. Gatto, {\PRD{81}{114031}{2010}}.}
\def\igormag{E.V. Gorbar, V.A. Miransky and I. Shovkovy,{\PRC{80}{032801(R)}{2009}};
ibid, arXiv:1009.1656[hep-ph].}
\def\miranski{V.P. Gusynin, V. Miranski and I. Shovkovy,{\PRL{73}{3499}{1994}};
{\PLB{349}{477}{1995}}; {\NPB{462}{249}{1996}}, E.J. Ferrer and V de la 
Incerra,{\PRL{102}{050402}{2009}}; {\NPB{824}{217}{2010}.}}
\def\providencia{D.P. Menezes, M. Benghi Pinto, S.S. Avancini and C. Providencia
,{\PRC{80}{065805}{2009}}; D.P. Menezes, M. Benghi Pinto, S.S. Avancini , A.P. Martinez
and C. Providencia, {\PRC{79}{035807}{2009}}}
\def\boomsma{J. K. Boomsma and D. Boer, {\PRD{81}{074005}{2010}}}
\def\somenath{D. Bandyopadhyaya, S. Chakrabarty and S. Pal, {\PRL{79}{2176}{1997}};
S. Chakrabarty, S. Mandal, {\PRC{75}{015805}{2007}}}
\def\klimenko{D. Ebert and K.G. Klimenko,{\NPA{728}{203}{2003}}}
\def\fukuwarringa{K. Fukushima and H. J. Warringa, {\PRL{100}{032007}{2008}}.}
\def\noornah{J. Noornah and I. Shovkovy,{\PRD{76}{105030}{2007}}.}
\def\armendirk{X.G. Huang, M. Huang, D.H. Rischke and A. Sedrakian, 
{\PRD{81}{045015}{2010}}.}
\def\metlitsky{M. A. Metlitsky and A. R. Zhitnitsky,{\PRD{72}{045011}{2005}}.}
\def\digal{T. Mandal, P. Jaikumar and S. Digal, arXiv:0912.1413 [nucl-th] .}
\def\dunc{ R. C. Duncan and C. Thompson, Astrophys. J. 392, L9 (1992).}
\def\duncc {C. Thompson and R. C. Duncan, Astrophys. J. 408, 194 (1993).}
 \def\dunccc{C. Thompson and R. C. Duncan, Mon. Not. R. Astron.  Soc. 275, 255 (1995).}
 \def\duncccc{C. Thompson and R. C. Duncan, Astrophys. J. 473, 322 (1996).}
 \def\kouvel{C. Kouveliotou et al., Astrophys. J. 510, L115 (1999).}
 \def\lat{C. Y. Cardall, M. Prakash, and J. M. Lattimer, Astrophys.  J. 554, 322 (2001).}
 \def\broder{A. E. Broderick, M. Prakash, and J. M. Lattimer, {\PLB{531}{167}{2002}}.}
 \def\lai{D. Lai and S. L. Shapiro, Astrophys. J. 383, 745 (1991).}
\def\dune{G.Baser, G. Dunne and D. Kharzeev, {\PRL{104}{232301}{2010}}.}
\def\frolov{I.E. Frolov, V. Ch. Zhukovsky and K.G. Klimenko, {\PRD{82}{076002}{2010}}.}
\def\nickel{D. Nickel,{\PRD{80}{074025}{2009}}.}
\def\gato{R. Gatto and M. Ruggieri, {\PRD{82}{054027}{2010}}.}
\def\iran{Sh. Fayazbakhsh and N. Sadhooghi, {\PRD{82}{045010}{2010}}.}
\def\hongmag{Deog Ki Hong, arXiv:1010.3923[hep-th].}
\def\bhamhmppn{B. Chatterjee, H. Mishra and A. Mishra; in preparation}

\end{document}